\documentclass[usenatbib,useAMS,usegraphicx]{mn2e}
\usepackage{comment}

\title[Recent star formation in high-redshift early-type galaxies]
    {The UV colours of high-redshift early-type galaxies: evidence for recent star
formation and stellar mass assembly over the last 8 billion years}

\author[S.~Kaviraj et al.]
{S. Kaviraj$^{1}$\thanks{E-mail: skaviraj@astro.ox.ac.uk}, S.
Khochfar$^{1}$, K. Schawinski$^{1}$, S. K. Yi$^{2}$, E.
Gawiser$^{3}$, J. Silk$^{1}$, \newauthor S. N. Virani$^{3}$, C. Cardamone$^{3}$, P. G. van Dokkum$^{3}$ and C. M. Urry$^{3}$\\
$^{1}$Department of Physics, University of Oxford, Keble Road,
  Oxford OX1 3RH, UK\\
$^{2}$Yonsei University, Centre for Space Astrophysics, Seoul
120749, Korea\\
$^{3}$Yale Center for Astronomy and Astrophysics, Yale University,
New Haven, CT 06520-8121}

\begin{document}
\date{5 September 2007}
\pagerange{\pageref{firstpage}--\pageref{lastpage}} \pubyear{2007}
\maketitle \label{firstpage}


\begin{abstract}
We combine deep optical and NIR ($UBVRIzJK$) photometry from the
Multiwavelength Survey by Yale-Chile (MUSYC) with redshifts from
the COMBO-17 survey to perform a large-scale study of the
rest-frame ultraviolet ($UV$) properties of 674 high-redshift
($0.5<z<1$) early-type galaxies, drawn from the Extended Chandra
Deep Field South (E-CDFS). Galaxy morphologies are determined
through visual inspection of Hubble Space Telescope (HST) images
taken from the GEMS survey. We harness the sensitivity of the $UV$
to young ($<1$ Gyrs old) stars to quantify the \emph{recent} star
formation history of early-type galaxies across a range of
luminosities ($-23.5<M(V)<-18$). Comparisons to simple stellar
populations forming at high redshift indicate that $\sim1.1$
percent of early-types in this sample are consistent with
\emph{purely} passive ageing since $z=2$ - this value drops to
$\sim0.24$ percent and $\sim0.15$ percent for $z=3$ and $z=5$
respectively. Parametrising the recent star formation (`RSF') in
terms of the mass fraction of stars less a Gyr old, we find that
the early-type population as a whole shows a typical RSF between 5
and 13 percent in the redshift range $0.5<z<1$. Early-types on the
broad $UV$ `red sequence', show RSF values less than 5 percent,
while the reddest early-types (which are also the most luminous)
are virtually quiescent with RSF values of $\sim1$ percent. In
contrast to their low-redshift ($z<0.1$) counterparts, the
high-redshift early-types in this sample show a pronounced
bimodality in the rest-frame $UV$-optical colour, with a minor but
significant peak centred on the blue cloud. Furthermore, star
formation in the most active early-types is a factor of 2 greater
at $z\sim0.7$ than in the local Universe. Given that evolved
sources of $UV$ flux (e.g. horizontal-branch stars) should be
absent at $z>0.5$, implying that the $UV$ is dominated by young
stars, we find compelling evidence that early-types of all
luminosities form stars over the lifetime of the Universe,
although the bulk of their star formation is already complete at
high redshift. This `tail-end' of star formation is measurable and
not negligible, with luminous ($-23<M(V)<-20.5$) early-types
potentially forming 10-15 percent of their mass since $z=1$, with
less luminous early-types ($M(V)>-20.5$) potentially forming 30-60
percent of their mass after $z=1$. This, in turn, implies that
intermediate-age stellar populations should be abundant in local
early-type galaxies, as expected in hierarchical cosmology.
\end{abstract}




\section{Introduction}
The star formation histories (SFHs) of early-type galaxies have
historically posed a significant challenge, with the exact
mechanism of their formation the subject of intense and
controversial debate in modern astrophysics. The classical
`monolithic collapse' hypothesis followed the model of
\cite{ELS62} for the formation of the Galaxy. Refined and
implemented by others \cite[e.g.][]{Larson74,Chiosi2002}, this
model postulated that stellar populations in early-type galaxies
form in short, highly efficient starbursts at high redshift
($z\gg1$) and evolve purely passively thereafter. The optical
properties of early-type populations and their strict obedience to
simple scaling relations are indeed remarkably consistent with
such a simple formation scenario. The small scatter in the
early-type `Fundamental Plane' \citep[e.g.][]{Jorg1996,Saglia1997}
and its apparent lack of evolution with look-back time
\citep[e.g.][]{Forbes1998,Peebles2002,Franx1993,Franx1995,VD1996},
the homogeneity \citep[e.g.][]{BLE92} and lack of redshift
evolution in their optical colours
\citep[e.g.][]{Bender1997,Ellis97,Stanford98,Gladders98,VD2000}
and evidence for short ($<1$ Gyr) star formation timescales in
these systems, deduced from the over-abundance of $\alpha$
elements \citep[e.g.][]{Thomas1999}, all indicate that the bulk of
the stellar population in early-type galaxies did indeed form at
high redshift ($z>2$).

While it reproduces the optical properties of early-type galaxies
remarkably well, a monolithic SFH does not to sit comfortably
within the currently accepted LCDM galaxy formation paradigm, in
which galaxy mass is thought to accumulate over the lifetime of
the Universe, through both quiescent and merger-driven star
formation. Following the seminal work of \citet{Toomre_mergers},
who showed that most galaxy collisions end in mergers and
postulated the formation of spheroidal systems as end-products of
such merging activity, the mechanics of galaxy interactions
\citep[e.g.][]{Barnes1992a,Barnes1992b,Hernquist1993} and their
link to the formation of early-type systems \citep[see e.g.][or
Bender (1996) for an observational perspective]{Barnes1996} has
been studied in considerable detail. Current semi-analytical
models of galaxy formation within the LCDM framework create
early-type galaxies through major mergers, where the mass ratio of
the merging progenitors is 3:1 or lower. The constituent stellar
mass of early-type systems is predicted to form both in mergers
and interactions and through quiescent star formation in their
progenitors over the lifetime of the
Universe\citep[e.g.][]{Cole2000,Hatton2003,Khochfar2003}.

While theoretical arguments may be compelling, the strongest
evidence for the role of interactions in shaping early-type galaxy
evolution and inducing coincident star formation comes from
observation. While individual examples of merger remnants can be
clearly seen in the local Universe (e.g. NGC 5128 \citep[][]{I98}
and IC 4200 \citep{Serra2006}), fine structure, which is
indicative of recent interactions is not uncommon in the nearby
early-type population. 40 percent of local ellipticals contain
dust lanes \citep{Sadler1985}, while over 75 percent contain
nuclear dust and by implication gas
\citep[e.g.][]{Tomita2000,Tran2001}. The gas is often
kinematically decoupled from the stars, indicating, at least in
part, an external origin, e.g. through the accretion of a gas rich
satellite \citep[e.g][]{sauron5}. Up to two-thirds of nearby
early-types contain shells and ripples
\citep[e.g.][]{Malin83,VD2005} and a significant fraction exhibit
kinematically decoupled cores \citep[e.g.][]{sauron2}, both of
which are evidence for interactions in the recent past
\citep[e.g.][]{Barnes1992b}. Moreover, fine structure and
disturbed morphologies often coincide with signatures of recent
star formation (e.g. blue broad-band colours), implying that such
events are frequently accompanied by detectable amounts of star
formation \citep[e.g.][]{Schweizer1990,Schweizer1992} even at low
redshift. While the detection of spatially resolved fine structure
is possible only for galaxies in our local neighbourhood,
signatures of recent star formation persist in the nearby galaxy
population out to modest redshifts
\citep[e.g.][]{Trager2000a,Trager2000b,Fukugita2004}.

While many of the recent efforts have centred on the nearby
Universe, deep ground and space-based imaging have increasingly
provided access to galaxy populations over the last ten billion
years of look-back time. A significant fraction ($\sim$30 percent)
of luminous early-type systems at high redshifts ($0.4<z<0.8$)
exhibit blue cores, indicative of merger-driven starbursts
triggered by the accretion of low-mass gas-rich companions
\citep{Menanteau2001a}. Not unexpectedly, such blue cores are
typically accompanied by emission and absorption lines
characteristic of recent star formation \citep{Ellis2001}. A
fundamental consequence of early-type evolution in the standard
model is the gradual loss of late-type progenitors over a Hubble
time \citep[e.g.][]{VD2001a,Kaviraj2005a}. Since late-types merge
to form early-types, the early-type fraction is expected to
decrease with increasing redshift, while the late-type fraction
should show a corresponding increase. Numerous observational
studies have detected this predicted evolution in the
morphological mix of the Universe. While $\sim$80 percent of
galaxies in the cores of local clusters have early-type morphology
\citep{Dressler80}, a higher fraction of spiral (blue) galaxies
have been reported in clusters at high redshift
\citep[e.g.][]{BO84,Dressler97,Couch98,VD2000,Margoniner2001,Andreon2004,Borch2006,Bundy2006},
along with increased rates of merger and interaction events
\citep[e.g.][]{Couch98,VD99}. The fraction of blue galaxies
evolves from a few percent in nearby ($z<0.1$) clusters to
$\sim30$ percent at intermediate redshifts ($z\sim0.5$) and
reaches $\sim70$ percent at high redshift $z\sim1$, although the
exact trend has some dependence on the cluster richness and
magnitude limit adopted by the study in question. This is
supported by recent results from large-scale surveys which suggest
that the mass density on the red sequence (which is dominated by
early-type systems) has doubled since $z=1$ (e.g. Bell et al.
2004; Faber et al. 2005). A significant body of observational
evidence thus indicates that the SFHs of at least some early-type
galaxies, and perhaps the early-type population as a whole,
deviate strongly from the expectations of the monolithic collapse
paradigm, both in terms of their structural evolution and the star
formation experienced by them over a Hubble time.

The vast majority of work on early-type galaxies in the past has
focussed on \emph{optical} spectro-photometric data. A significant
drawback of optical photometry is its lack of sensitivity to
moderate amounts of recent star formation. While red optical
colours do imply a high-redshift formation epoch for the bulk of
the stars in early-type galaxies, the optical spectrum remains
largely unaffected by the minority of stellar mass that forms at
low and intermediate redshift. As a result it is virtually
impossible to resolve early-type SFHs over the last 8 billion
years ($0<z<1$), where the predictions of the two competing models
diverge the most.

A first step towards resolving the SFHs over this timescale is to
quantify the \emph{recent star formation} (RSF) in early-type
systems at $z\sim0$. RSF can be efficiently traced using the
rest-frame ultraviolet ($UV$) spectrum, which is sensitive to
young, massive main sequence stars with ages less than $\sim$1
Gyr. The $UV$ remains largely unaffected by the `age-metallicity
degeneracy' which plagues optical studies \citep{Worthey1994},
allowing it to maintain its age sensitivity across a large range
in metallicity and making it an ideal photometric indicator of RSF
(Kaviraj et al. 2006c). Using a large sample of early-type
galaxies detected by the GALEX $UV$ space telescope, Kaviraj et
al. (2006b) have recently shown that, contrary to the expectations
of classical models \citep[e.g][]{BLE92}, nearby early-type
galaxies show widespread RSF - at least 30 percent show blue
$UV$-optical colours which cannot be achieved through the $UV$
flux from old horizontal-branch (HB) stars \citep{Yi97,Yi99} alone
and require some RSF to be present in these systems. In fact, up
to 90 percent of early-types in this sample have photometry that
is consistent with RSF within the observational errors. Most
importantly, Kaviraj et al. (2006b) have demonstrated that the
early-type SFHs predicted by semi-analytical models in the LCDM
framework provide excellent quantitative agreement with the
observed $UV$ photometry from GALEX and that monolithic models
(where RSF can be driven solely by recycled gas from stellar mass
loss) cannot reproduce the observed $UV$ colours under reasonable
assumptions for the chemical enrichment or star formation
histories.

This study extends these results to high redshift by exploiting
deep optical photometry to trace the rest-frame $UV$ properties of
early-type galaxies in the redshift range $0.5<z<1$. We harness
the sensitivity of the $UV$ to young stars to quantify the recent
star formation history of early-types at high redshift. We combine
our results with similar studies in the local Universe (Kaviraj et
al. 2006b) to estimate the star formation history of the
early-type population between $z=0$ and $z=1$ (i.e. the last 8
billion years of look-back time) and test their conformity to
scenarios for their formation, especially the classical
`monolithic collapse' hypothesis.

We note that the $UV$ properties of the E-CDFS galaxy population
have previously been explored by \citet[][W05
hereafter]{Wolf2005}, who looked at the contributions to the $UV$
luminosity density from various morphological classes of galaxies
in the redshift range $0.65<z<0.75$. The study presented here
differs from W05 in several ways. While W05 studied the $UV$ flux
around $2800 \AA$, our emphasis is on the $UV$ spectrum around
$2300 \AA$ which is more sensitive to recent star formation.
Furthermore, we deal exclusively with the early-type population
over a larger redshift range $0.5<z<1$ and explicitly quantify the
recent star formation in each galaxy in our sample. Nevertheless,
W05 forms an important benchmark for the results presented here
and we draw parallels betweeb the two studies throughout the
course of this work.


\section{Early-type galaxies in the Extended Chandra Deep Field South}
\subsection{The data}
Numerous ongoing optical surveys are now deep and wide enough to
provide access to the rest-frame $UV$ flux of entire galaxy
populations at redshifts beyond $\sim 0.5$. The Extended Chandra
Deep Field South (E-CDFS), in particular, contains a plethora of
deep surveys which provide panchromatic data over a large range in
redshift, spanning the X-rays through to the infrared wavelengths.
Three recent optical surveys in E-CDFS, MUSYC (Gawiser et al.
2006)\footnote{http://www.astro.yale.edu/MUSYC/}, COMBO-17
\citep{Wolf2004}\footnote{http://www.mpia.de/COMBO/combo\_index.html}
and GEMS (Rix et al.
2004)\footnote{http://www.mpia.de/GEMS/home.htm}, provide the full
suite of deep optical photometry, redshifts and morphologies
required to accurately study the early-type population in the
rest-frame $UV$ at high redshift.

The MUSYC survey provides deep $UBVRIz'JK$ imaging of E-CDFS, to
AB depths of $U,B,V,R=26.5$ and $J,K=22.5$. The MUSYC broad-band
($UVBRI$) data combines COMBO-17 imaging with those from the ESO
Deep Public Survey \citep{Arnouts2001}. COMBO-17 provides 17
filter photometry, including 12 medium-band filters, and accurate
photometric redshifts due to its large filter set. The redshift
accuracies are $\sim1$ percent for $R<21$, $\sim2$ percent for
$R\sim22$ and better than 10 percent for $R<24$. GEMS provides $V$
and $z$-band HST-ACS images of galaxies in E-CDFS out to $z\sim1$,
from which morphologies can be robustly deduced, due to the superb
angular resolution of the ACS camera. The GEMS survey traces at
least the rest-frame $B$-band within our target redshift range
($0.5<z<1$), so that morphological K-corrections are minimal
within this redshift range.

In this study, we combine deep broad-band photometry from MUSYC
and photometric redshifts from COMBO-17 with morphologies from
GEMS to study the early-type population in E-CDFS in the
rest-frame $UV$ within the redshift range $0.5<z<1$. The lower
limit of our target redshift range ($z\sim0.5$) is determined by
the epoch at which the optical $U$-band ($\sim3600 \AA$) traces
the rest-frame near-$UV$ ($\sim2300\AA$). The upper limit
($z\sim1$) is driven primarily by the fiducial depth of GEMS, but
also by the epoch to which redshifts can be reliably determined
from COMBO-17 and the need to avoid large morphological
K-corrections. In addition, we adopt a magnitude limit of $R=24$,
because COMBO-17 redshifts are robust down to this limit and the
GEMS images are bright enough to perform reasonably accurate
morphological classification.

We note that, although MUSYC and COMBO-17 offer similar broad-band
filter sets, MUSYC is significantly deeper than COMBO-17 in its
broad-band imaging - e.g. the MUSYC $U$-band is 2 magnitudes
deeper than its COMBO-17 counterpart. It is the depth of this
imaging, particularly in the $U$ and $B$-band filters, that makes
it possible to trace the rest-frame near-$UV$ accurately. Since
errors in the optical photometry propagate through the analysis
into the estimated rest-frame photometry, a robust determination
of the rest-frame colours and their associated parameters requires
deep optical imaging. We also note that, wherever possible, we
substitute COMBO-17 photometric redshifts with spectroscopic
redshifts from the VVDS survey (Le F{\`e}vre et al. 2004). In
addition, for galaxies that fall within the GOODS (Giavalisco et
al. 2004) field in E-CDFS, we use GOODS imaging instead of GEMS
since the GOODS pointings are deeper.

Finally, we note that (Type 1) AGN are removed using the COMBO-17
`QSO' flag and by removing objects that are detected in the E-CDFS
X-ray point-source catalog of \citet{Virani2006}. The fraction of
early-types in this X-ray catalog is small - only 15 out of the
674 objects that are morphologically classified as early-type (see
the next section) have detections in \citet{Virani2006}.


\subsection{Morphological classifications}
The advent of large-scale space and ground-based surveys in a wide
variety of wavelengths is giving us unprecedented access to
statistically large populations of galaxies across a variety of
redshifts and environments. However, the size of the galaxy
samples poses a challenge when morphological classification is
necessary. Early-type galaxies can be selected using automated
pipelines that isolate objects on the basis of their
two-dimensional light distributions. For example, Bernardi et al.
(2000) have constructed a catalog of $\sim$ 9000 low-redshift
early-type galaxies, selected using a combination of SDSS pipeline
parameters. This catalog contains galaxies with a high $i$-band
concentration index ($r_{50}/r_{90}> 2.5$) and in which a
deVaucouleurs fit to the surface brightness profile is
significantly more likely than an exponential fit. Visual
inspection of this catalog indicates that, while such automated
prescriptions are efficient at selecting a reasonably robust
early-type galaxy sample, they do admit a significant fraction of
contaminants (e.g. face-on disks, edge-on spirals and late-type
galaxies with weak spiral features).

In this study we simply opt for visual inspection of the
individual HST images of MUSYC galaxies in E-CDFS. Although it is
arguably subjective and logistically difficult to implement on
large samples of galaxies, visual inspection minimises the
contamination in the early-type galaxy sample studied in this
paper. For the sample studied in this paper, the superb angular
resolution ($\sim0.05"$) of the HST facilitates morphological
classification, since small-scale structure is better resolved
than ground-based images which typically have resolutions of
$\sim1"$. At the imaging depth of the GEMS and GOODS surveys (from
which the HST imaging of the MUSYC galaxies is derived) and within
the magnitude limit adopted in this study, morphological
classification of early-type galaxies should be `safer' than that
performed using standard SDSS images in Kaviraj et al. (2006b),
who studied the $UV$ properties of a sample of low-redshift
($z<0.1$) early-type galaxies drawn, through visual inspection,
from the SDSS.

In this study, we split our parent galaxy sample (comprised of
4498 objects) into three main morphological classes. The first are
`relaxed early-type' galaxies which show an unperturbed spheroidal
morphology with no detectable signs of spiral features. We refer
to this class simply as `early-type' hereafter. The second are
`possible early-types' which appear early-type but whose
spheroidal morphology is uncertain. As our analysis indicates
below, it is likely that these galaxies are indeed contaminants,
as the distribution of their properties does not follow that of
the (relaxed) early-type population. Separating these objects is
clearly important and makes our `early-type' classification safer.
The third class, which we label `late-type', includes all objects
which do not fall into the above categories. We also identify a
fourth category of `disturbed galaxies' (e.g. tidal tails and
double cores) which are potential \emph{progenitors} of early-type
galaxies and which are the subject of a forthcoming paper.

\begin{figure}
$\begin{array}{c}
\includegraphics[width=3.3in]{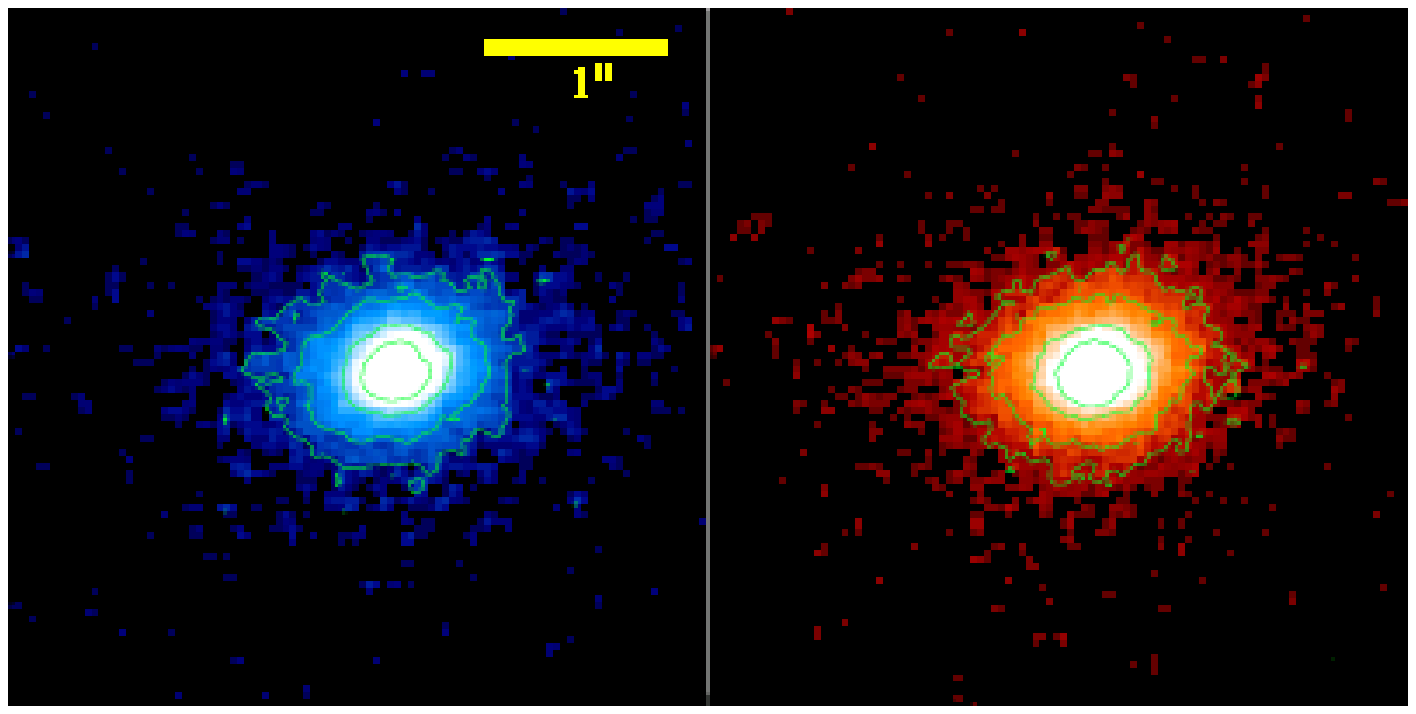}\\
\includegraphics[width=3.3in]{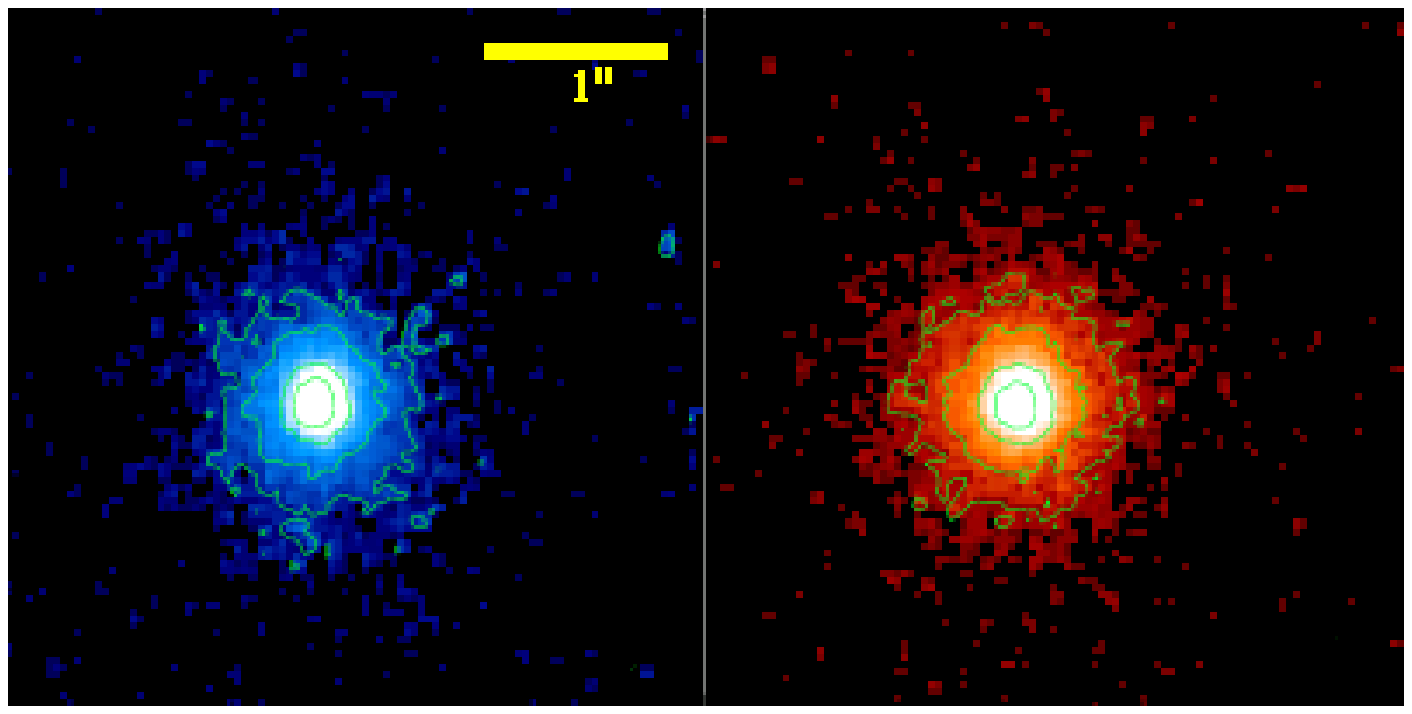}
\end{array}$
\caption{Example GEMS images of galaxies classified as
`early-type' in this study. The left-hand column shows the
$V$-band image (rest-frame $U$-band), while the right-hand column
shows the $z$-band image (rest-frame $V$-band). The top panel
shows an early-type at $z=0.596$, while the bottom panel shows an
early-type at $z=0.720$. The redshifts are photometric and taken
from the COMBO-17 survey. 1" corresponds to $\sim7$ kpc at
$z\sim0.7$.} \label{fig:etype_examples}
\end{figure}

\begin{figure}
$\begin{array}{c}
\includegraphics[width=3.3in]{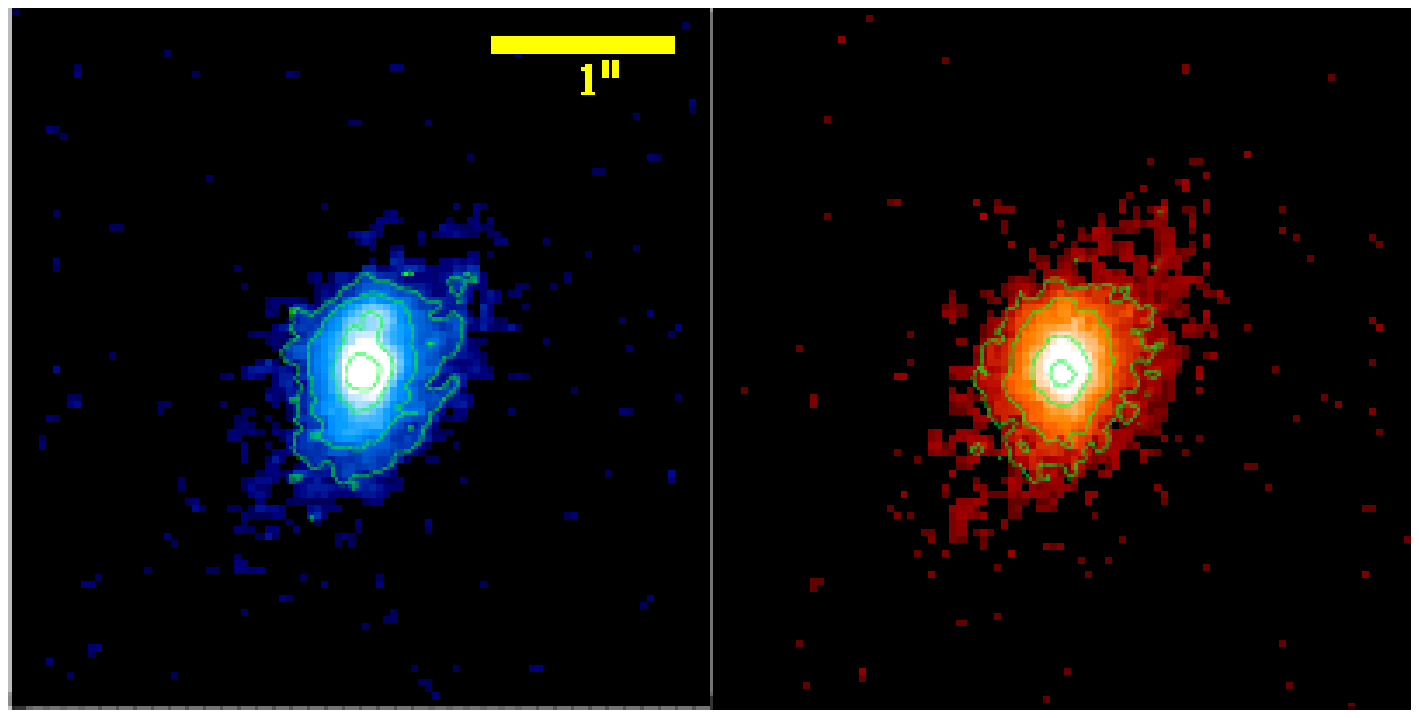}\\
\includegraphics[width=3.3in]{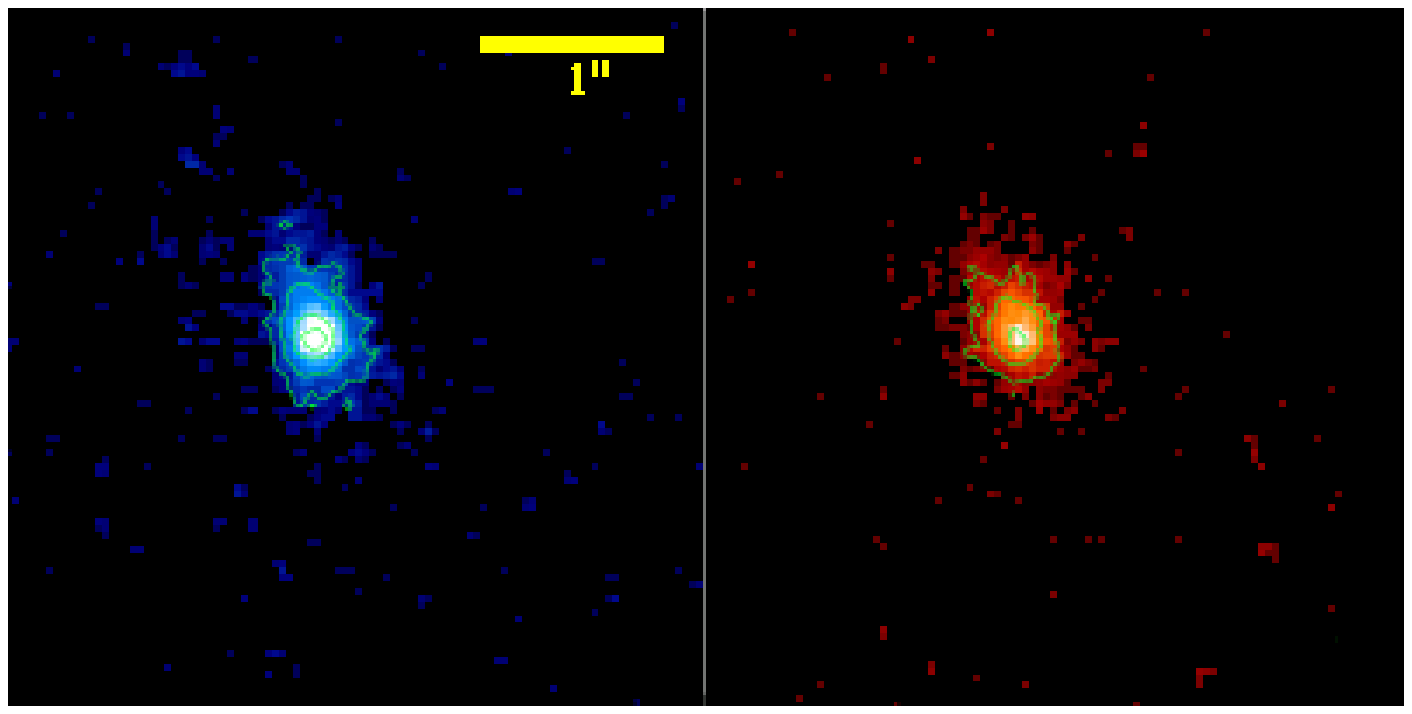}
\end{array}$
\caption{Example GEMS images of galaxies classified as `possible
early-type' in this study. The left-hand column shows the $V$-band
image (rest-frame $U$-band), while the right-hand column shows the
$z$-band image (rest-frame $V$-band). The top panel shows a
possible early-type at $z=0.637$, while the bottom panel shows a
possible early-type at $z=0.607$. The redshifts are photometric
and taken from the COMBO-17 survey. 1" corresponds to $\sim7$ kpc
at $z\sim0.7$.} \label{fig:petype_examples}
\end{figure}

Figures \ref{fig:etype_examples} and \ref{fig:petype_examples}
show example images of the `early-type' and `possible early-type'
categories. We show GEMS images of each galaxy in both the
$V$-band (rest-frame $U$-band; left-hand column) and the $z$-band
(rest-frame $V$-band; right-hand column). The morphological
classification was carefully performed by SK, using both bands and
checked against composite images released by the GEMS
collaboration. The process was repeated five times and the
fraction of galaxies in the `possible early-type' category became
stable between the third and fourth passes. This process yields
674 early-type galaxies and 100 `possible early- types', out of a
parent sample of 4498 objects.


To check the quality of the visual inspection we compare our
classification results (for galaxies in the redshift range
$0.65<z<0.75$) to those of W05, who also performed visual
inspection of their E-CDFS galaxy sample with a similar magnitude
cut of $m(R)<24$. Unfortunately W05 do not provide a breakdown of
the number of objects in each of their morphological classes so
the comparison is performed by counting up the relevant number of
objects in the upper left panel of Figure 5 and dividing by the
total number of objects in the W05 sample. Since early-types on
the red sequence can be easily identified (the red sequence is
dominated by early-types in all studies) we simply compare the
fraction of early-types in the blue cloud selected by the two
studies. We find that blue ($M_{280}-M_V<1$) early-types in W05
comprise $\sim9.3$ percent of their parent galaxy population. In
comparison blue early-types in our study, selected using the
criterion $(NUV-r)<4$, comprise $6.8$ percent of the population.
This figure rises to $8.4$ percent if we also consider blue
galaxies (again using $(NUV-r)<4$) in the possible early-type
category. Since the colour cut employed to separate the red
sequence from the blue cloud is somewhat arbitrary, we do not find
the small difference in the blue early-type fractions particularly
compelling and are satisfied that the visual inspection in these
two studies correspond well. In fact, it is likely that the
classification performed in this study is slightly more
conservative than that used by W05.


\subsection{Estimation of rest-frame photometry and derived parameters}
We estimate the rest-frame photometry and star formation histories
(SFHs) of individual galaxies by comparing the multi-wavelength
photometry of each observed MUSYC object with synthetic galaxy
populations, generated in the framework of the standard model.
Synthetic populations are generated using the semi-analytical
model of \citet{Khochfar2003}. Cosmological parameters used for
the background cosmology are taken from the three-year WMAP
observations (Spergel et al. 2007): $\Omega_m = 0.241$,
$\Omega_{\Lambda} = 0.759$, $h=0.732$, $\sigma_8 = 0.761$. Merger
trees are calculated using the extended Press-Schechter formalism
with a mass resolution of $5 \times 10^9 M_{\odot}$. The model
reproduces merger statistics and galaxy properties accurate down
to $1 \times 10^{10} M_{\odot}$ \citep{Khochfar2001}. Baryonic
physics is implemented as described in \citet{Khochfar2005} and
\citet{Khochfar2006} and star formation histories include the
self-consistently calculated contributions of all progenitor
galaxies. AGN feedback is implemented according to the
phenomenological approach described in \citet{Schawinski2006}, in
which gas cooling is halted in galaxies hosting black holes more
massive than a critical black hole mass.

A library of synthetic photometry is constructed by combining each
of $\sim$15,000 model galaxies with a single metallicity in the
range $0.1Z_{\odot}$ to $2.5Z_{\odot}$ and a measure of the dust
content parametrised by a value of $E(B-V)$ in the range 0 to 0.5.
Photometric predictions are generated by combining each model
combination with the stellar models of \cite{Yi2003} and the MUSYC
filter transmission curves convolved with the correct atmospheric
conditions. Each library thus contains $\sim$750,000 model
galaxies and their associated predicted photometry in the MUSYC
filter set.

Since our observed dataset spans a large redshift range
($0.5<z<0.1$), synthetic libraries are generated at redshift
intervals of $\delta z = 0.02$. Each MUSYC object is studied using
the library which is closest to it in redshift. A fine redshift
grid minimises `K-correction-like' errors in the estimated
rest-frame colours that are induced by the redshift offset between
the observed object and the library used to analyse it. Each
observed galaxy is compared to every model in the library and the
likelihood ($\exp{-\chi^2/2}$) of each model extracted using the
value of $\chi^2$, computed in the standard way. We estimate
rest-frame colours (e.g. $U-V$, $NUV-r$) by taking the average
colour of all models which yield `good fits' ($\chi^2_R<2$). We
take the standard deviation of these model colours as a measure of
the error in the rest-frame photometry. Finally, we construct an
`average SFH' for each observed galaxy, by combining the SFHs of
all models in the library weighted by their individual
likelihoods. The recent star formation (RSF), which is the focus
of this study and which we define as the \emph{mass fraction of
stars formed within the last Gyr in the rest-frame of the galaxy},
is calculated from this average SFH. Note that the $NUV$ passband
employed in this study corresponds to the GALEX $NUV$ filter which
covers the spectral range $1750\AA-2750\AA$ and has an effective
wavelength of $2271\AA$.



\section{Derived rest-frame $UV$-optical colours of the E-CDFS galaxy population}

\subsection{Sources of $UV$ flux in early-type galaxies}
Analysis of the rest-frame $UV$ properties of early-type galaxies
in the local Universe ($0<z<0.2$) is complicated by the fact that
their $UV$ spectrum may contain contributions from both young and
old stellar populations. Core helium burning stars on the
horizontal branch (HB), thought to be the primary cause of the `UV
upturn' phenomenon in massive elliptical galaxies
\citep{Yi97,Yi99}, emit efficiently in the $UV$ spectral ranges.
Thus the potential `contamination' of the $UV$ spectrum from old
HB stars must be taken into account before estimating the
contribution from young stellar populations, complicating the
robust detection of recent star formation in early-type systems.

The onset of the HB typically takes 9-10 Gyrs
\citep[e.g.][]{Yi99}, depending on the metallicity of the
population. Thus, a key benefit of the analysis presented here is
that, given the lower redshift limit ($z=0.5$) employed in this
study, HB stars should not have time to develop and contribute to
the $UV$ spectrum\footnote{Assuming formation redshifts of 3 and
5, the maximum age of stellar populations are $\sim6.3$ and
$\sim7.3$ Gyrs respectively}. The lack of evolved $UV$-emitting
stellar populations therefore makes our conclusions more robust
than can be achieved by corresponding studies in the local
Universe (e.g. Kaviraj et al. 2006b).


\subsection{Rest-frame colour-magnitude relations}
\begin{figure}
$\begin{array}{c}
\includegraphics[width=3.5in]{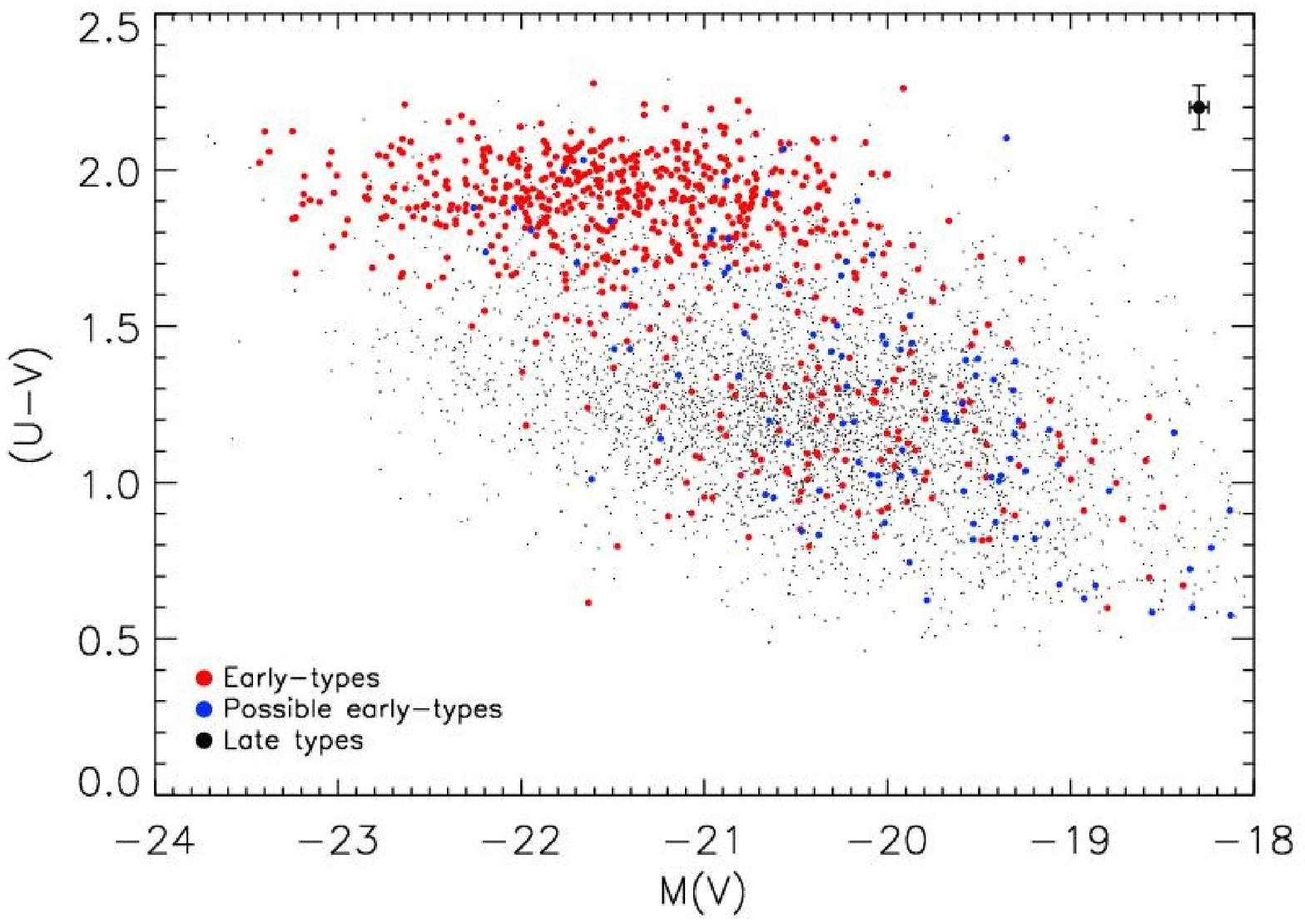}\\
\includegraphics[width=3.5in]{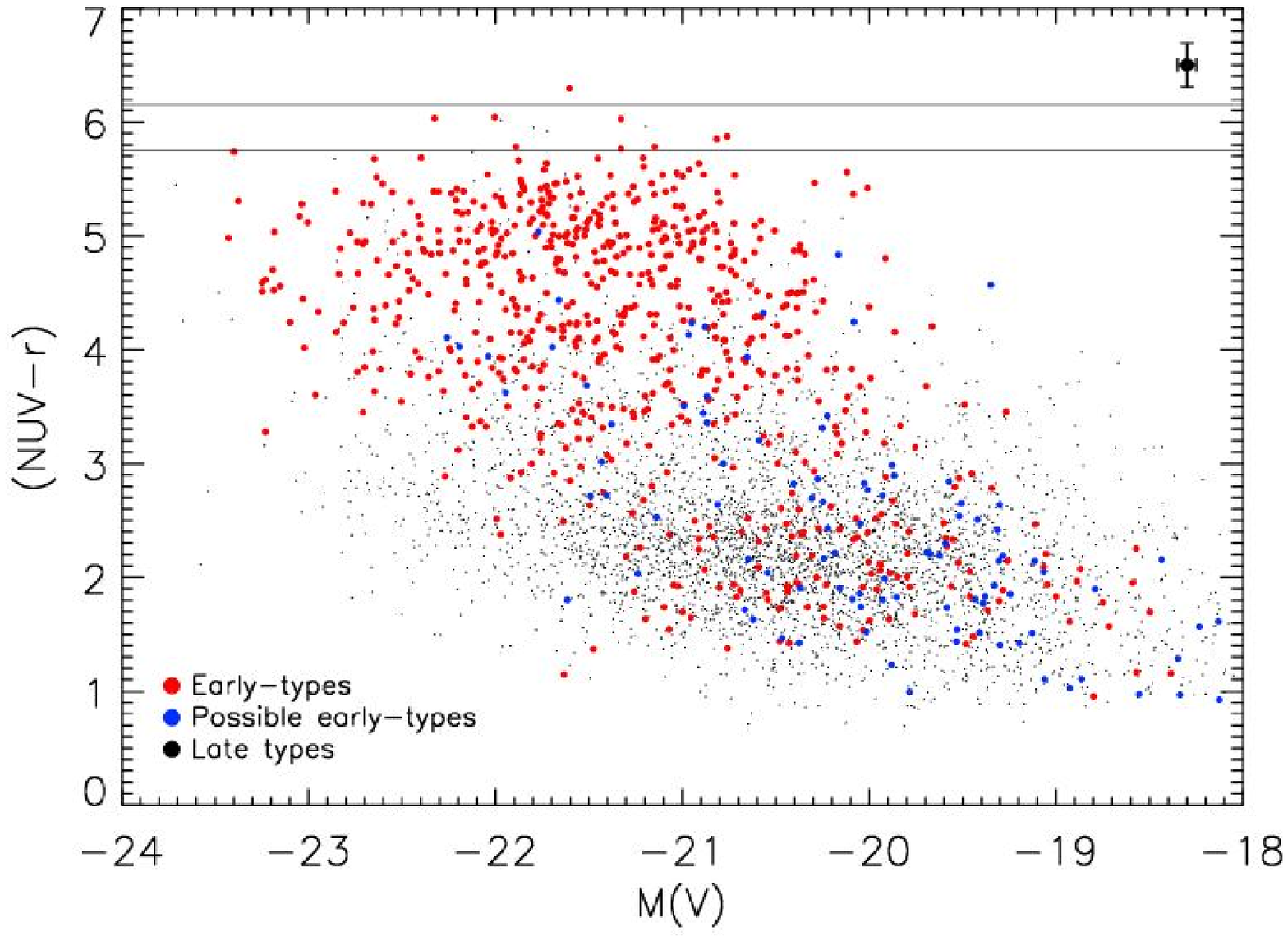}
\end{array}$
\caption{TOP PANEL: The rest-frame $(U-V)$ CMR of the galaxy
population in the E-CDFS in the redshift range $0.5<z<1$. BOTTOM
PANEL: The rest-frame $(NUV-r)$ CMR of the galaxy population in
the E-CDFS. The $NUV$ passband is taken from the GALEX filterset
and the $r$-band filter is taken from the SDSS. The filters are
chosen to facilitate comparison with the results of Kaviraj et al.
(2006b) at low redshift. The solid horizontal lines indicate the
expected position of a dustless simple stellar population which
forms at $z=3$ with solar metallicity and is observed at $z=0.5$
(upper line) and $z=1$ (lower line).} \label{fig:cmr}
\end{figure}

\begin{figure}
$\begin{array}{c}
\includegraphics[width=3.5in]{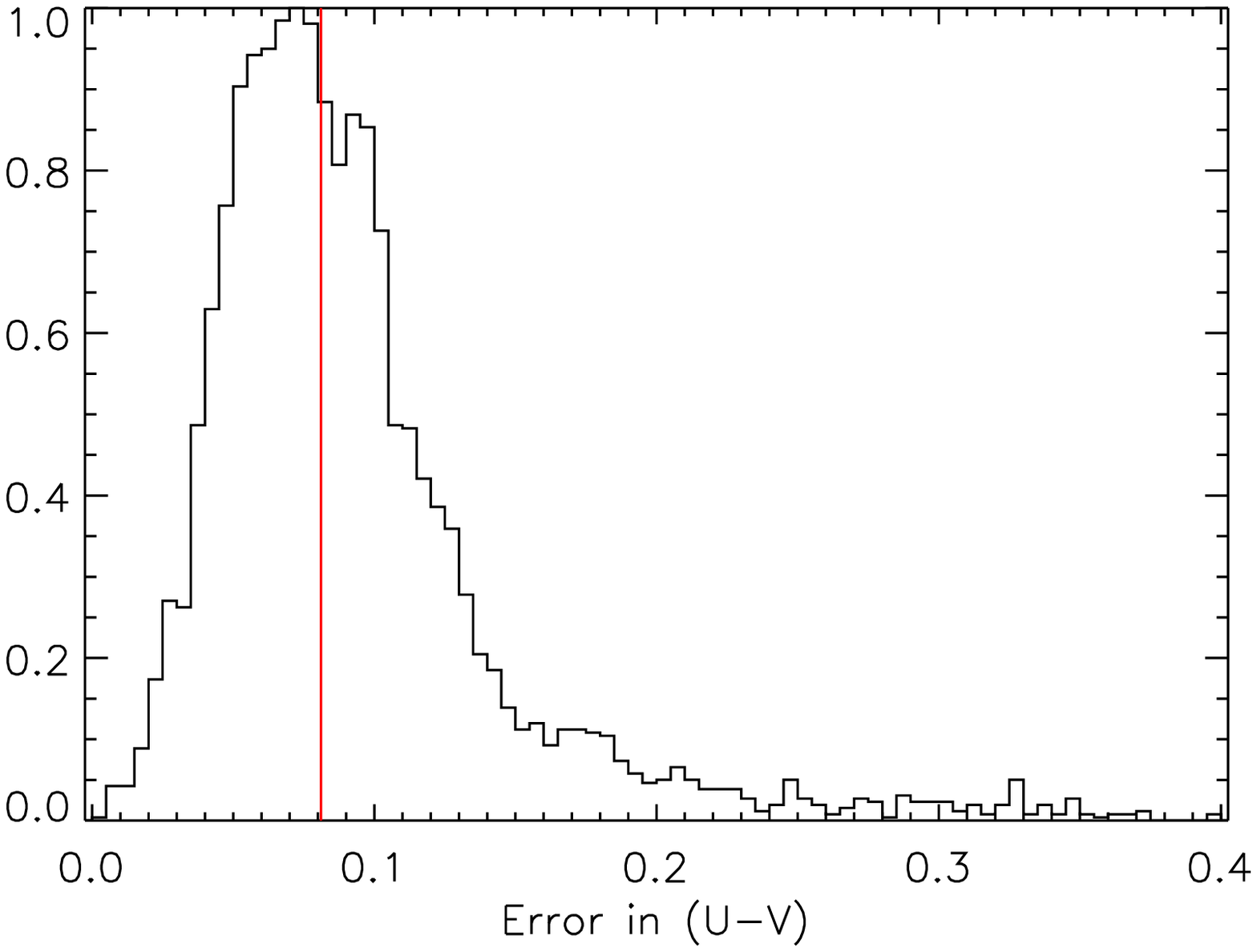}\\
\includegraphics[width=3.5in]{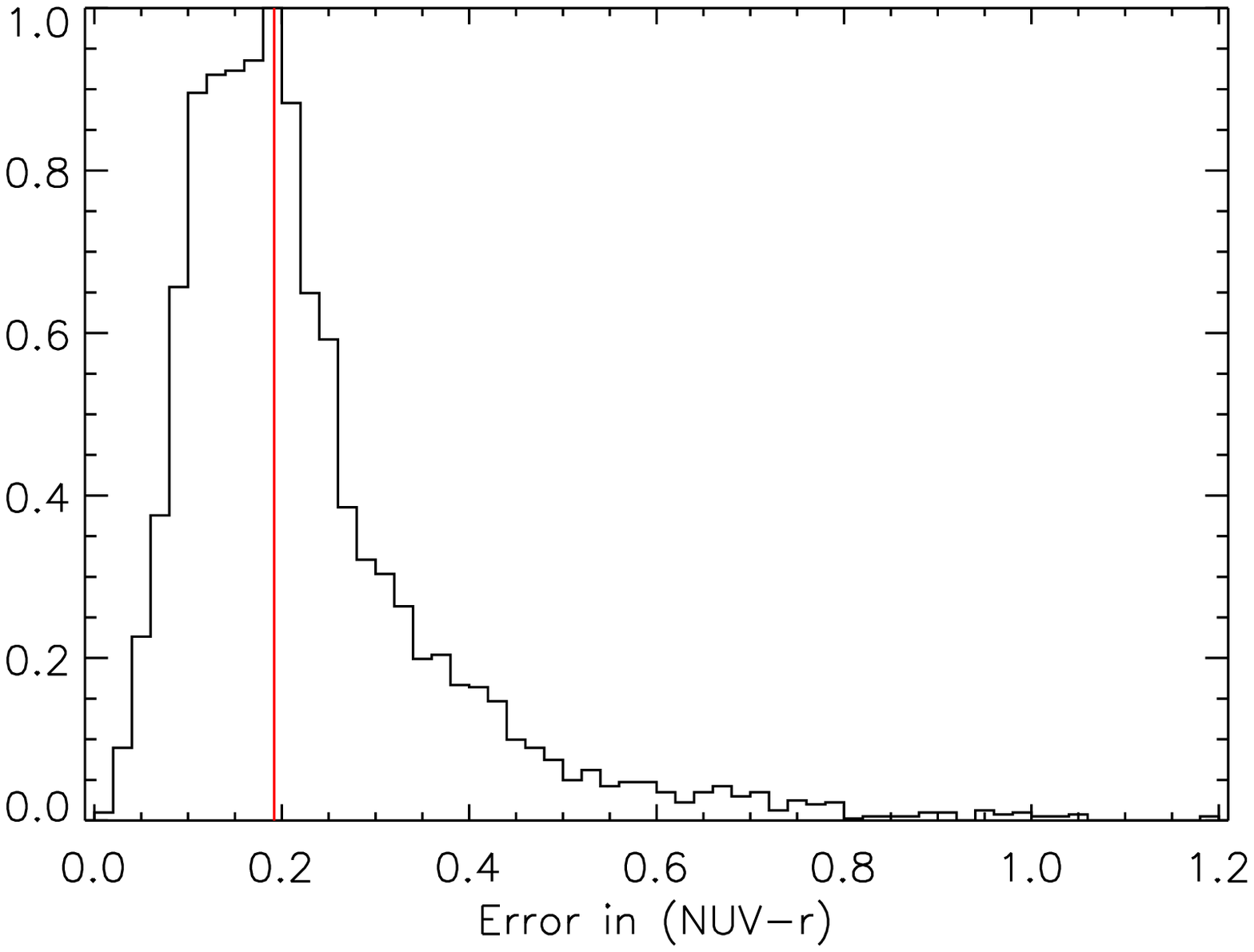}
\end{array}$
\caption{Errors in the derived rest-frame $U-V$ (top) and $NUV-r$
(bottom) colours of the galaxy population in E-CDFS (shown in
Figure \ref{fig:cmr}). Median uncertainties are indicated using
the solid red lines. The median uncertainty in $(U-V)$ is
$\sim0.08$ while the median uncertainty in $(NUV-r)$ is
$\sim0.2$.} \label{fig:colour_errors}
\end{figure}

\begin{figure}
$\begin{array}{c}
\includegraphics[width=3.3in]{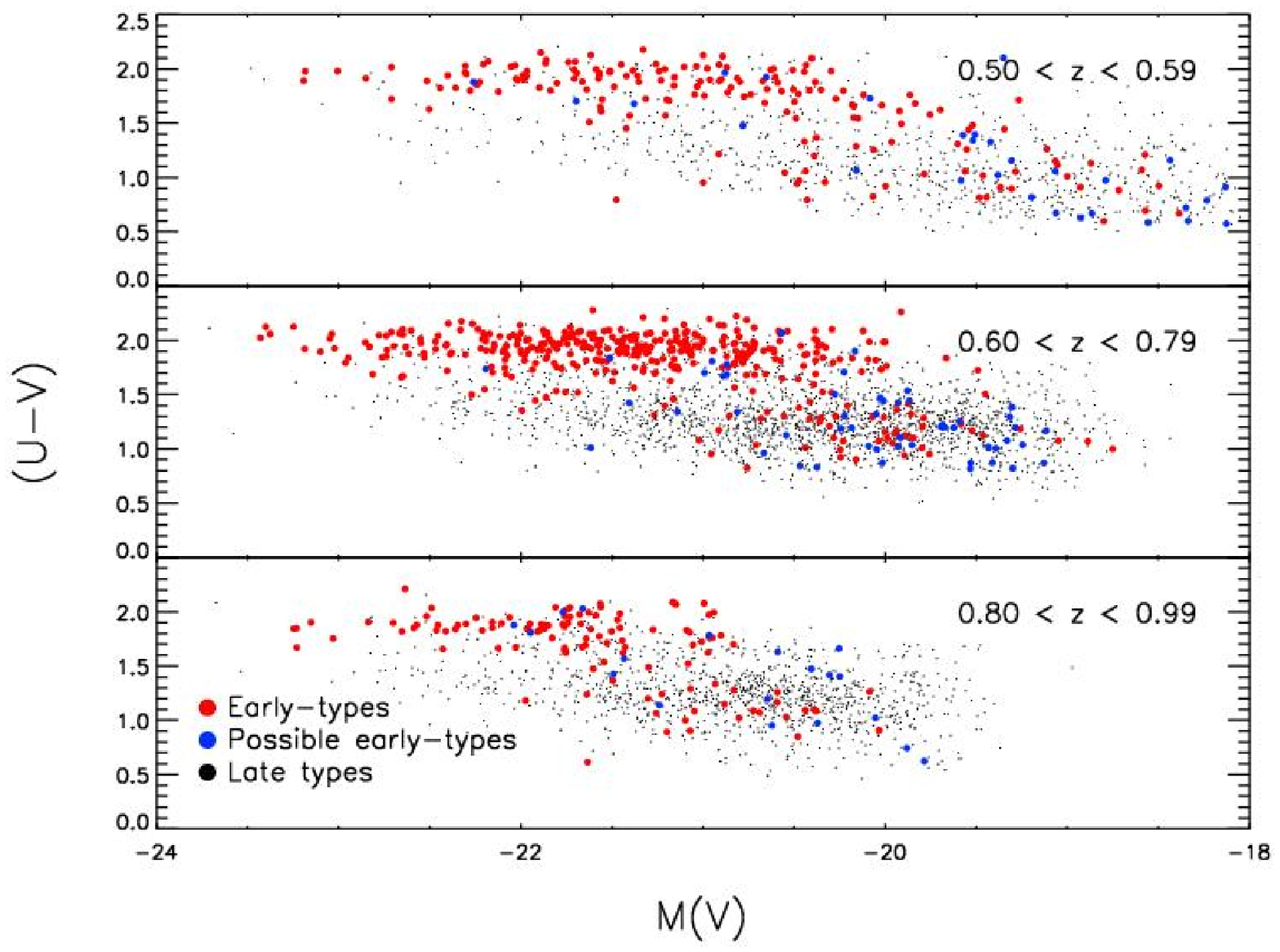}\\
\includegraphics[width=3.3in]{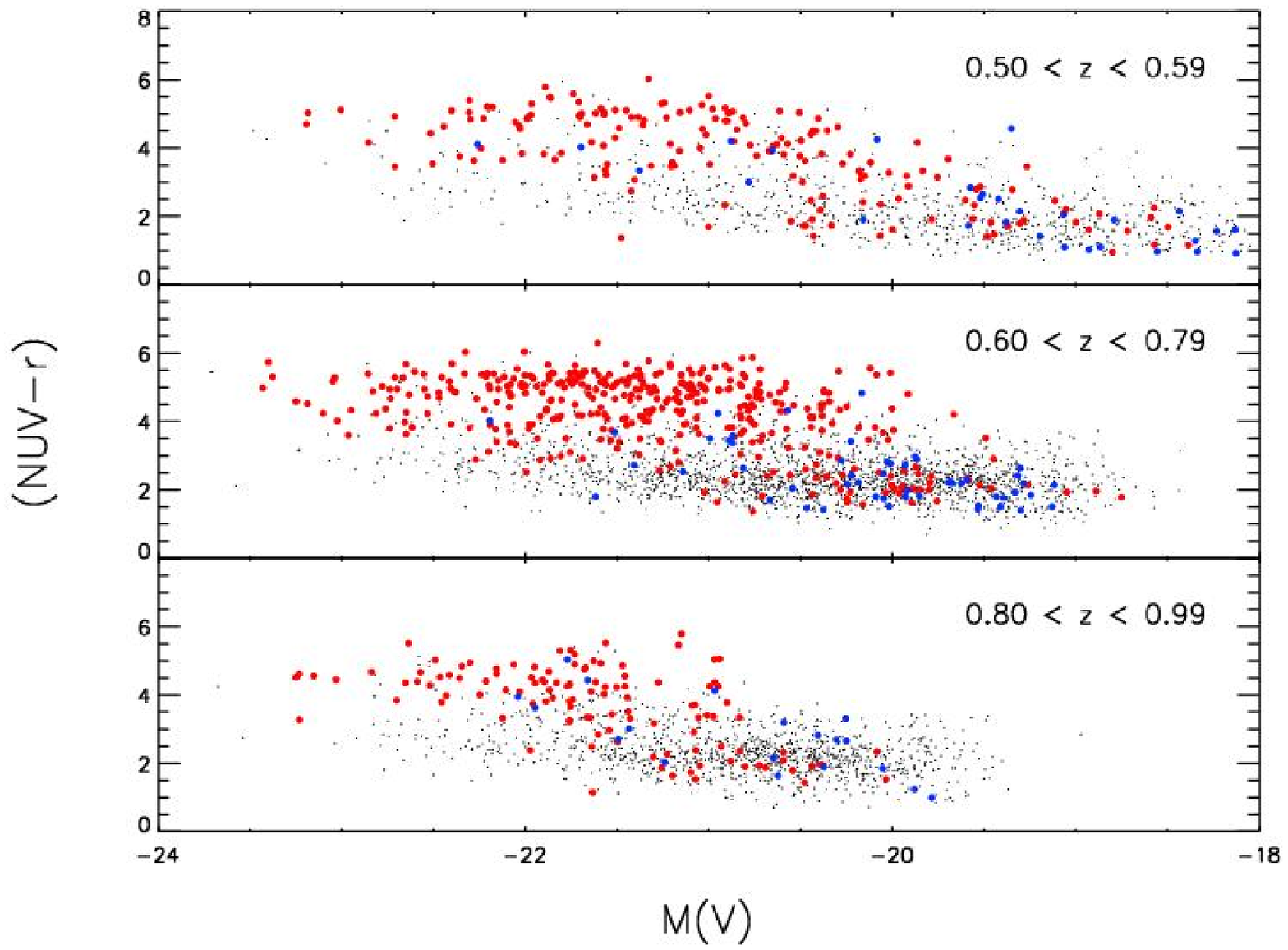}
\end{array}$
\caption{TOP PANEL: The rest-frame $(U-V)$ CMR as a function of
redshift. BOTTOM PANEL: The rest-frame $(NUV-r)$ CMR as a function
of redshift.} \label{fig:cmr_binned}
\end{figure}

We begin by presenting the estimated rest-frame photometry for the
galaxy population in E-CDFS within the redshift range $0.5<z<1$.
Figure \ref{fig:cmr} shows the rest-frame $(U-V)$ (top) and
$(NUV-r)$ (bottom) CMRs of the galaxy population in the E-CDFS in
the redshift range $0.5<z<1$. Red circles indicate early-type
galaxies and green circles represent possible early-types. The
small black dots indicate late-type systems. The $(U-V)$ colours
are in excellent agreement with the COMBO-17 results of
\citet{Bell2004} for the E-CDFS. Note that magnitudes given here
are in the AB system while COMBO-17 reports magnitudes in the
VEGAMAG system. The conversion between the two filter systems is
$(U-V)_{MUSYC} \sim (U-V)_{COMBO17} + 0.75$. The $NUV$ passband is
taken from the GALEX filterset and the $r$-band filter is taken
from the SDSS. These filters are chosen to facilitate comparison
with the results of Kaviraj et al. (2006b) at low redshift. Figure
\ref{fig:colour_errors} presents the distribution of one-sigma
errors for the derived rest-frame $(U-V)$ and $(NUV-r)$ colours
shown in Figure \ref{fig:cmr}. Median uncertainties are indicated
using the solid red lines. The median uncertainties in $(U-V)$ is
$\sim0.08$, while the median uncertainty in $(NUV-r)$ is $\sim0.2$
\footnote{For comparison, the observational errors in the GALEX
(Martin et al. 2005) $UV$ photometry of low-redshift galaxies is
$\sim0.25$ magnitudes.}. In Figure \ref{fig:cmr_binned} we split
the rest-frame $(U-V)$ and $(NUV-r)$ CMRs into three redshifts
bins to show their evolution with look-back time.

Similar to the early-type population at low redshift (see Figure 4
in Kaviraj et al. 2006b), the high-redshift early-types in this
study show a broad red sequence, which spans almost three
magnitudes in the $(NUV-r)$ colour. The origin of this broadness
is the high sensitivity of the $UV$ to young stars - the $UV$
photometry is able to resolve small differences in the recent SFH
resulting in the broad distribution of colours on the red
sequence. Conversely, the insensitivity of optical colours to such
low levels of RSF results in the significantly tighter optical red
sequence that is apparent in the top panel of Figure
\ref{fig:cmr}. We note that a similar result was found by W05, who
found a red sequence spanning 1.5 mag in the rest-frame ($2800-V$)
colour (see their Figure 5). The fact that the ($2800-V$) red
sequence is narrower is expected, since it is not as sensitive to
recent star formation as its $(NUV-r)$ counterpart.

\subsection{Comparisons to the monolithic collapse hypothesis}
It is instructive to first compare the expectations of the
classical monolithic scenario, in which galaxies evolve
\emph{completely passively} since high redshift, to the rest-frame
CMRs of the E-CDFS population. For example, the \citet{Yi2003}
models predict that a (dustless) simple stellar population (SSP)
of solar metallicity which forms at $z\sim3$\footnote{The choice
of $z=3$ is motivated by optical studies
\citep[e.g.][]{BLE92,Bower98} which have `age-dated' cluster
early-type populations using optical photometry. The SSP age
estimates from such studies consistently indicate a formation
redshift greater than $z=2$.}, will have $(NUV-r) \sim 6.15$ and
$(NUV-r) \sim 5.75$ at $z\sim0.5$ and $z\sim1$ respectively. As an
independent check we calculate the position of this dustless $z=3$
SSP using the \citet{BC2003} stellar models and find that its
corresponding position remains virtually unchanged at $(NUV-r)
\sim 6.12$ and $(NUV-r) \sim 5.71$ for $z\sim0.5$ and $z\sim1$
respectively.

Inspection of Figure \ref{fig:cmr} indicates that early-types in
general are incompatible with such a simple SFH, since virtually
all early-types in this sample (even those in the red sequence)
lie blueward of $(NUV-r)\sim5.75$. We make this comparison more
robust by comparing the $(NUV-r)$ colour of each galaxy in our
sample to SSPs forming at high redshift ($z=2,3,5$). We apply the
median metallicity of the galaxy in question to the SSP but note
that our results remain virtually unchanged if we simply assume
solar metallicity throughout. A galaxy that has an $(NUV-r)$
colour redder, within errors, than that of an SSP forming at
$z=z_f$ is clearly consistent with being passively evolving since
$z=z_f$. Thus, by comparing the $(NUV-r)$ colour of each galaxy to
SSPs at $z_f=2,3$ and 5 we can estimate the fraction of
early-types which could be considered to be ageing passively (i.e.
evolving `monolithically') since these formation redshifts. We
find that within the sample studied here, $\sim1.1$ percent of
early-types are consistent with purely passive ageing since $z=2$.
This value drops to $\sim0.24$ percent and $\sim0.15$ percent for
$z=3$ and $z=5$ respectively. This result includes the individual
errors in $(NUV-r)$ for each galaxy and is therefore robust. We
therefore find that, based on this SSP comparison, virtually none
of the early-type galaxies in this sample exhibit photometric
properties consistent with purely passive ageing since $z=2$.

While the SSP criterion used above to estimate the fraction of
potentially passively evolving galaxies is justifiable and driven
by previous observational studies, it is reasonable to ask whether
the choice of a single metallicity affects our interpretation, as
real galaxies - even those that might evolve monolithically - must
contain a metallicity distribution (MF) in their stellar
populations. We note first that MFs derived from semi-analytic
models of early-type galaxy formation, which are potentially more
realistic, are very sharply peaked at solar or super-solar values,
depending on the mass of the early-type in question (Nagashima \&
Okamoto 2006; Seong-Hee Kim, priv. comm.), implying that the
single metallicity assumption is a reasonable one.

In a more relevant study, Kaviraj et al. (2006c, see their Section
6) explored hypothetical $(NUV-r)$ colours of `monolithically
evolving' scenarios using chemical enrichment models. Monolithic
galaxies were constructed by requiring high star-formation
efficiencies at high redshift, resulting in red optical colours
and high values of alpha-enhancement as observed in the local
early-type population \citep[e.g.][]{Thomas1999}. Note that, by
construction, these models contain metalicity distributions.
Comparison of the $(NUV-r)$ colours at $z\sim0.5$ and $z\sim1$ of
such monolithic scenarios indicates that they are consistent with
the SSP colours we have used in our analysis. Depending on the
strength of the galactic wind allowed in these models the
$(NUV-r)$ colours can, in fact, be redder (by as much as $\sim0.2$
mag) than a solar metallicity SSP, especially in deep potential
wells where metals are more efficiently retained. Therefore, the
fractions of potentially quiescent galaxies derived above, using
comparisons to SSPs, should be reasonably robust.

\begin{figure}
\includegraphics[width=3.5in]{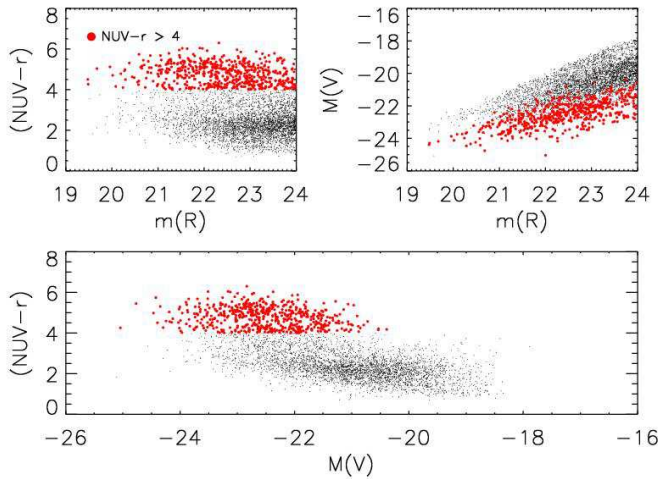}
\caption{Although red ($NUV-r>4$) galaxies are well-sampled in
apparent $R$-band magnitude (top left), the K-corrections for red
objects are larger (top right), resulting in these systems
shifting further to the \emph{left} in the rest-frame CMR (bottom
panel). This induces a lack of red galaxies in the rest-frame CMR
as shown in bottom panel. To populate the faint red end in the
rest-frame CMR requires galaxies which have $m(R)<24$. Hence the
lack of red galaxies is caused indirectly by our magnitude limit
($m(R)=24$). Recall, however, that our magnitude limit is chosen
to ensure reliability of the redshift estimates and morphological
classifications.} \label{fig:redlack}
\end{figure}


\subsection{The lack of faint red galaxies}
The rest-frame CMRs in Figure \ref{fig:cmr} show a distinct lack
of \emph{red} galaxies at the low luminosity end. We explain this
briefly using Figure \ref{fig:redlack} which indicates that, while
red ($NUV-r>4$) galaxies are well-sampled in apparent $r$-band
magnitude (top left), the K-corrections for red objects are larger
(top right), resulting in these systems shifting further to the
\emph{left} in the rest-frame CMR. This induces a lack of red
galaxies at the low luminosity end of the rest-frame CMR, as shown
in bottom panel. To populate this low luminosity end therefore
requires galaxies which have $m(R)>24$. Hence, the lack of red
galaxies is indirectly caused by our magnitude limit of $m(R)=24$.
Recall, however, that this magnitude limit is chosen to ensure the
reliability of the redshift estimates and robustness of the
morphological classifications and thus such low luminosity red
galaxies are beyond the scope of this study.


\section{Quantifying the recent star formation in the E-CDFS galaxy population}
The main focus of this study is to quantify the recent star
formation (RSF) in the high redshift ($0.5<z<1$) early-type
population. Recall that we define the RSF as the mass fraction of
stars that form in a galaxy, within the last 1 Gyr of look-back
time in its rest frame. The choice of timescale is driven by the
fact that the rest-frame $UV$ is most sensitive to stars less than
$\sim1$ Gyr old.

\begin{figure}
$\begin{array}{c}
\includegraphics[width=3.5in]{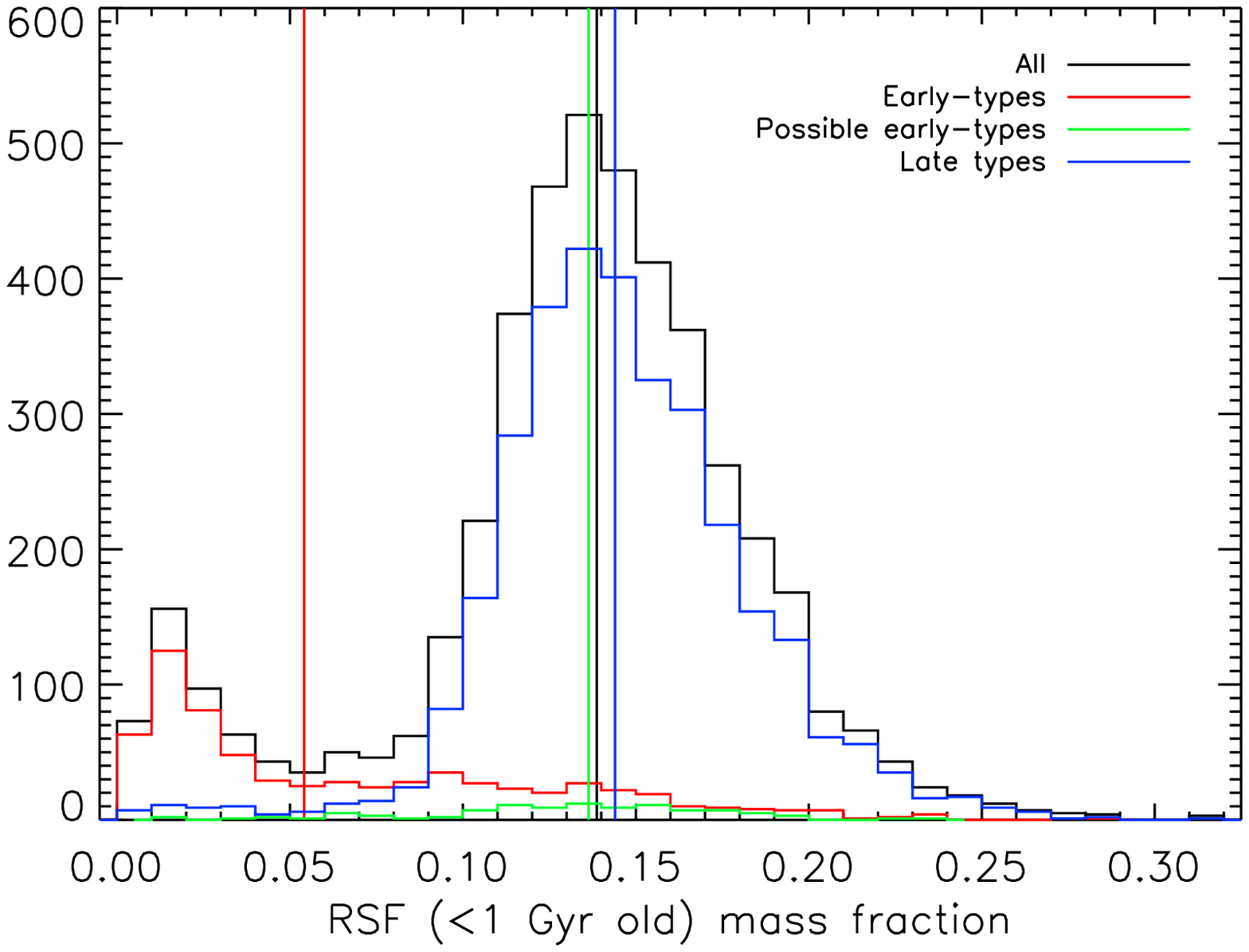}\\
\includegraphics[width=3.5in]{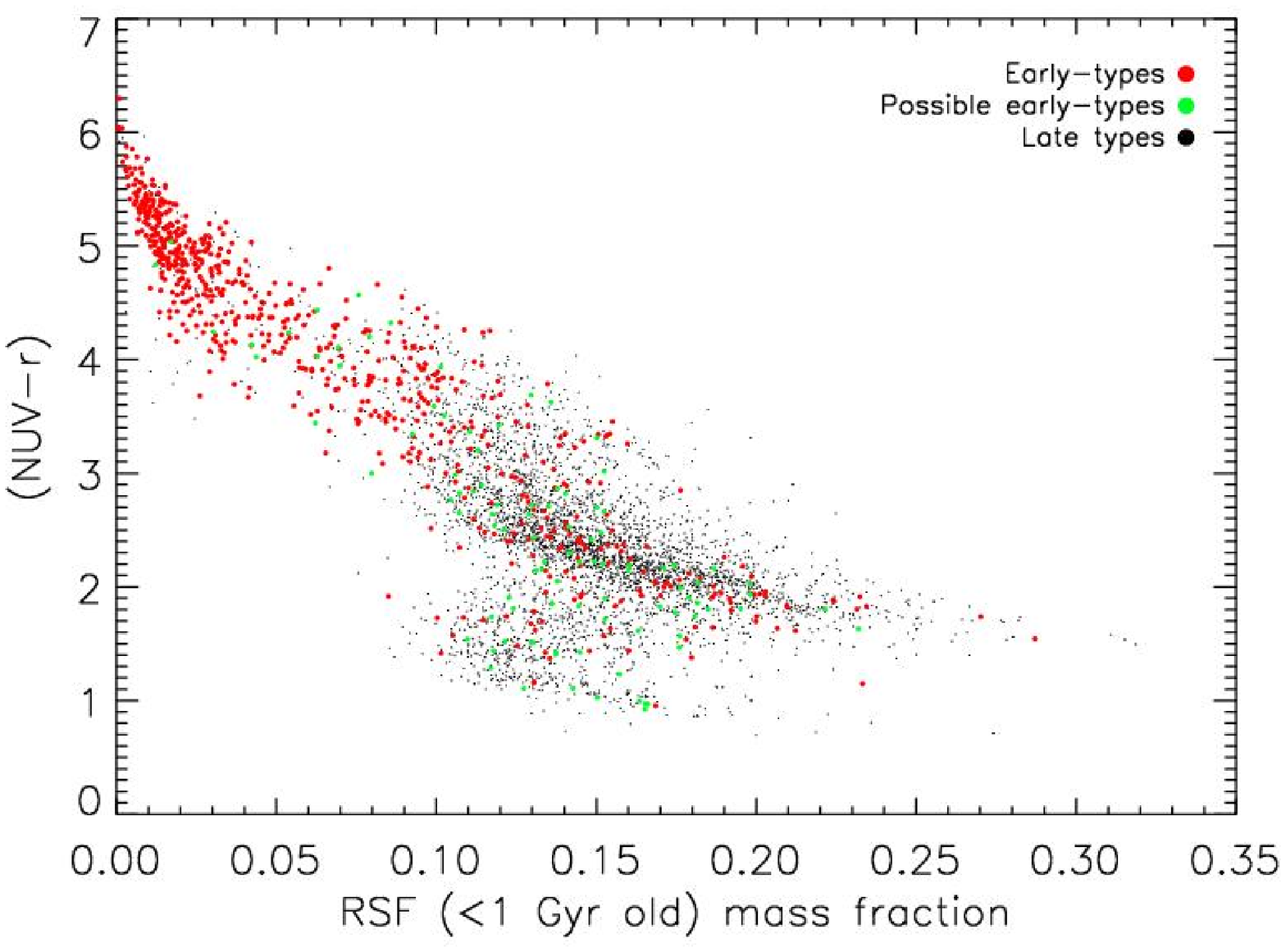}
\end{array}$
\caption{TOP PANEL: Histogram of recent star formation (RSF) for
galaxies in the E-CDFS population. Recall that RSF is defined as
the mass fraction of stars formed within the last Gyr of look-back
time in the rest-frame of the galaxy. Solid vertical lines
indicate the median values of the distributions for different
morphologies. The RSF is bimodal with the small peak at very low
RSF values completely dominated by (luminous) early-type galaxies
(see below). BOTTOM PANEL: The rest-frame $(NUV-r)$ colour plotted
against the RSF. A good separation exists between morphologies in
this parameter space, with early-types (red) dominating the clump
of galaxies with very low RSF values. Late type galaxies span a
wide range in RSF values from $\sim5$ percent up to $\sim30$
percent of the total galaxy mass. The broadening of the relation
at high RSF values is caused by widely disparate recent SFHs
within the galaxy population (see the text in Section 4 for
further discussion).} \label{fig:nuvr_rsf}
\end{figure}

Figure \ref{fig:nuvr_rsf} (top panel) shows the histogram of RSF
values for all galaxies in the E-CDFS population. The RSF
distribution is bimodal. The population as a whole shows a median
RSF of $\sim14$ percent (marked using the solid black line in
Figure \ref{fig:nuvr_rsf}) while the maximum value is $\sim30$
percent. Late types (blue) span a wide range of RSF values from
$\sim5$ percent up to $\sim30$ percent of the galaxy mass. Median
values of RSF (indicated by the vertical lines in the inset) are
$\sim5.5$ percent for early-types (red), $\sim13.8$ percent for
the `possible' early-type population (green) and $\sim14.4$
percent for late-type galaxies (black). The late-type population
typically shows three times the RSF of their early-type
counterparts. We also note that the bulk of the early-type
population, which preferentially populate the red sequence
($NUV-r>4$), have RSF values lower than the median value of $5.5$
percent (the modal value is $\sim2$ percent) - the slightly high
median is caused by the long tail to high RSF values.

The bottom panel of Figure \ref{fig:nuvr_rsf} shows the rest-frame
$(NUV-r)$ colour plotted against the RSF and split by galaxy
morphology. The small peak at low RSF values, apparent in the RSF
distribution (top panel), is completely dominated by early-type
galaxies (red points in the bottom panel). A robust separation
exists between morphologies in this parameter space, with
early-types overwhelmingly dominating the clump of galaxies with
very low ($<5$ percent) RSF values.

It should be noted that the value of the RSF mass fraction does
not give any indication of the \emph{profile} of the recent SFH.
Galaxies with similar RSF mass fractions can have widely disparate
$(NUV-r)$ colours, depending on the exact shape of their recent
SFHs. For example, a galaxy experiencing a burst of star formation
nearly 1 Gyr in the past will be redder in the $(NUV-r)$ colour
than a galaxy experiencing a burst of similar intensity which
takes place within the last 100 Myrs. Hence, while a correlation
clearly must exist between RSF and the $(NUV-r)$ colour, the
variation in the recent SFHs induces a `broadening' in the
$(NUV-r)$ vs. RSF relation, which is apparent in Figure
\ref{fig:nuvr_rsf}. A spread in dust and metallicity values may
also contribute to this broadening, but their contribution is
minor compared to that caused by the variation in recent SFHs
within the galaxy population. In particular, we find some galaxies
in the $(NUV-r)$ vs. RSF parameter space, with RSF values between
10 and 20 percent, which are displaced from the main locus of
galaxies. These objects show bluer $(NUV-r)$ colours for the same
values of RSF as galaxies on the main locus because they have
encountered a very recent period of star formation, probably as a
result of an interaction event, within the last 0.5 to 1 Gyr. We
also observe that, not unexpectedly, the correlation between
$(NUV-r)$ colour and RSF becomes tighter for low values of RSF,
since very small starbursts, even at small look-back times, are
not able to significantly perturb the $(NUV-r)$ colour from that
due to a largely old population.

\begin{figure}
\includegraphics[width=3.5in]{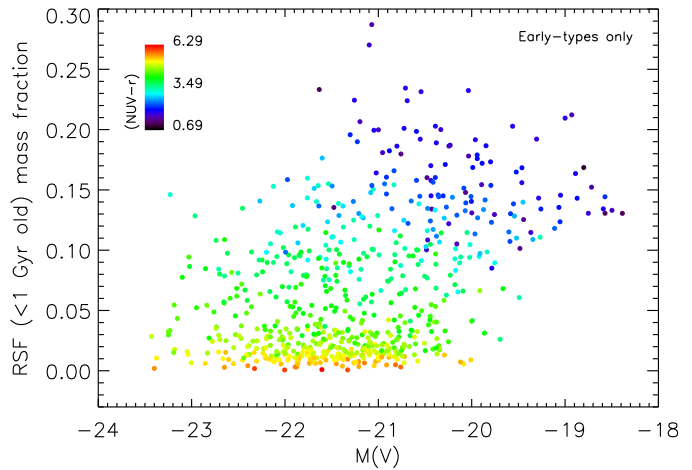}
\caption{RSF as a function of the absolute $V$-band luminosity of
the early-type galaxies, with galaxies colour-coded according to
their rest-frame $(NUV-r)$ colour. Note that the other
morphological classes are omitted from this figure for clarity.}
\label{fig:rsf_mv}
\end{figure}

In Figure \ref{fig:rsf_mv} we show the RSF as a function of the
absolute $V$-band luminosity of the early-type galaxies, with
galaxies colour-coded according to their $(NUV-r)$ colour. Note
that the other morphological classes are omitted from this figure
for clarity. A well-populated locus with low RSF values (less than
5 percent) exists across a wide range in luminosities
($-23.5<M(V)<-20$). The high RSF tail, apparent in the top panel
of Figure \ref{fig:nuvr_rsf}, is not restricted only to faint
early-type galaxies. A minority ($\sim10$ percent) of luminous
early-types ($-23.5<M(V)<-21$) exhibit high values of RSF, greater
than 10 percent. Galaxies fainter than $M(V)=-20$ typically have
very high RSF values (between 10 and 25 percent). Finally, there
is a broad trend of increasing RSF towards fainter galaxies, a
manifestation of the `downsizing' phenomenon recently studied in
the literature \citep{Cowie1996,Bundy2006}.

\begin{figure}
\includegraphics[width=3.5in]{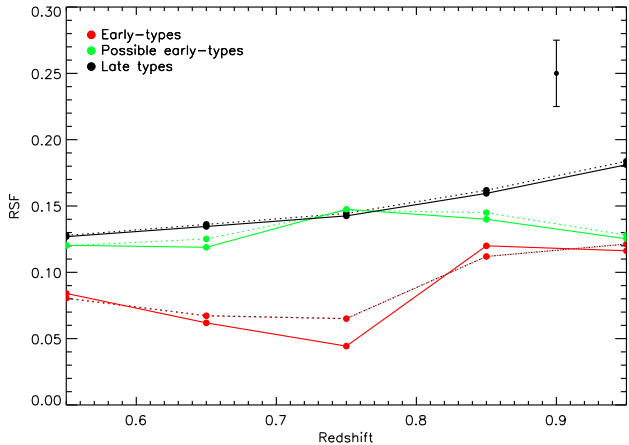}
\caption{Evolution of RSF with redshift for all the morphological
classes. Solid lines in this figure represent median values of RSF
(in keeping with results presented earlier) while the dotted lines
represent mean values.} Note that most of the redshifts in this
study are photometric (see text for
details).\label{fig:rsfevolution}
\end{figure}

In Figure \ref{fig:rsfevolution} we show the evolution of RSF with
redshift (across all the morphological classes). Solid lines in
this figure represent median values of RSF (in keeping with
results presented earlier) while the dotted lines represent mean
values. Within the errors, there is a slight indication that the
RSF increases with redshift, from $\sim7$ percent at $z=0.5$ to
$\sim13$ percent at $z=1$. The typical uncertainty in the RSF
values is $\sim2.5$ percent.

\subsection{Comparison to the local Universe}
\begin{figure}
\includegraphics[width=3.5in]{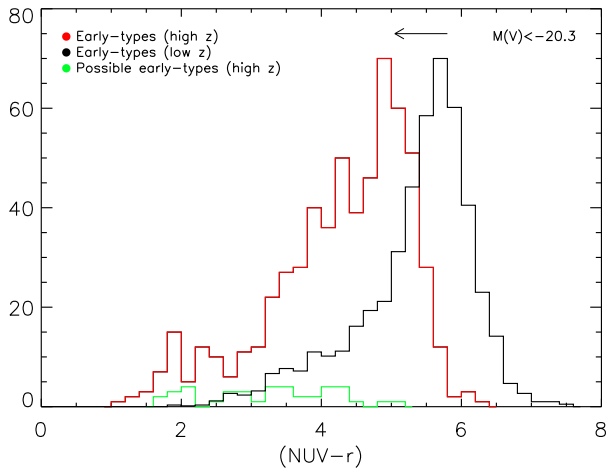}
\caption{Comparison of the high-redshift $(NUV-r)$ colours in this
study (solid red line) to their counterparts at low redshift
(solid black line) from Kaviraj et al. (2006b). The histograms are
normalised to the same peak value to facilitate comparison. Also
shown are the histograms for possible early-types (green) and
disturbed galaxies which could be progenitors of early-types at
lower redshift (blue). The arrow indicates the change in the
$(NUV-r)$ colour of a solar metallicity SSP population which forms
at $z=3$ from a redshift of $z\sim0.7$ to $z\sim0.1$, which
approximately represent the median redshifts in both studies.}
\label{fig:comparetolowz}
\end{figure}

Figure \ref{fig:comparetolowz} compares the high-redshift
$(NUV-r)$ colours in this study (solid red line) to their
counterparts at low redshift (solid black line) from Kaviraj et
al. (2006b). In contrast to the low-redshift early-type population
studied by Kaviraj et al. (2006b), high-redshift early-types show
a pronounced bi-modality around $(NUV-r)\sim3$, with the blue peak
centred on the blue cloud shown in Figure \ref{fig:cmr}. The peak
of the high-redshift $(NUV-r)$ distribution shows a relative shift
from its low-redshift counterpart (shown using the arrow in Figure
\ref{fig:comparetolowz}) which corresponds to the passive ageing
of a simple stellar population forming at $z=3$. This indicates
that the bulk of stellar population in early-type galaxies (on the
red sequence) is overwhelmingly old.


The blue peak contains $\sim15$ percent of the early-type
population, with an average RSF of $\sim11$ percent. In
comparison, the bluest 15 percent of the low-redshift early-type
population, studied by Kaviraj et al. (2006b), has an average RSF
of $\sim6$ percent. The intensity of (recent) star formation in
the most active early-types has therefore halved between
$z\sim0.7$ (which it median redshift  of this study) and
present-day.


\subsection{The dust-age degeneracy in $UV$ colours - robustness of the RSF estimation}
As mentioned before, the $UV$ spectrum is largely unaffected by
the age-metallicity degeneracy (Kaviraj et al. 2006c) and
therefore the age estimation remains robust within the metallicity
range considered in this study ($0.1Z_{\odot}$ to $2.5Z_{\odot}$).
However, the $UV$ is far more sensitive to the presence of dust
than the optical spectrum. The GALEX $NUV$ filter for example, on
which the rest-frame analysis in this paper in based, is three
times more sensitive to dust than the optical $V$-band filter,
assuming standard dust extinction laws
\cite[e.g.][]{Cardelli1989,Calzetti2000}. It is thus possible to
`redden' the $UV$ colour by adding arbitrary amounts of dust to
the system. Therefore, as an independent check of the robustness
of the parameter estimation we demonstrate that the derived SFH
parameters for red early-type galaxies do not involve significant
amounts of dust, since these galaxies should be red due to a lack
of young stars rather than due to heavy internal extinction.

\begin{figure}
$\begin{array}{c}
\includegraphics[width=3.5in]{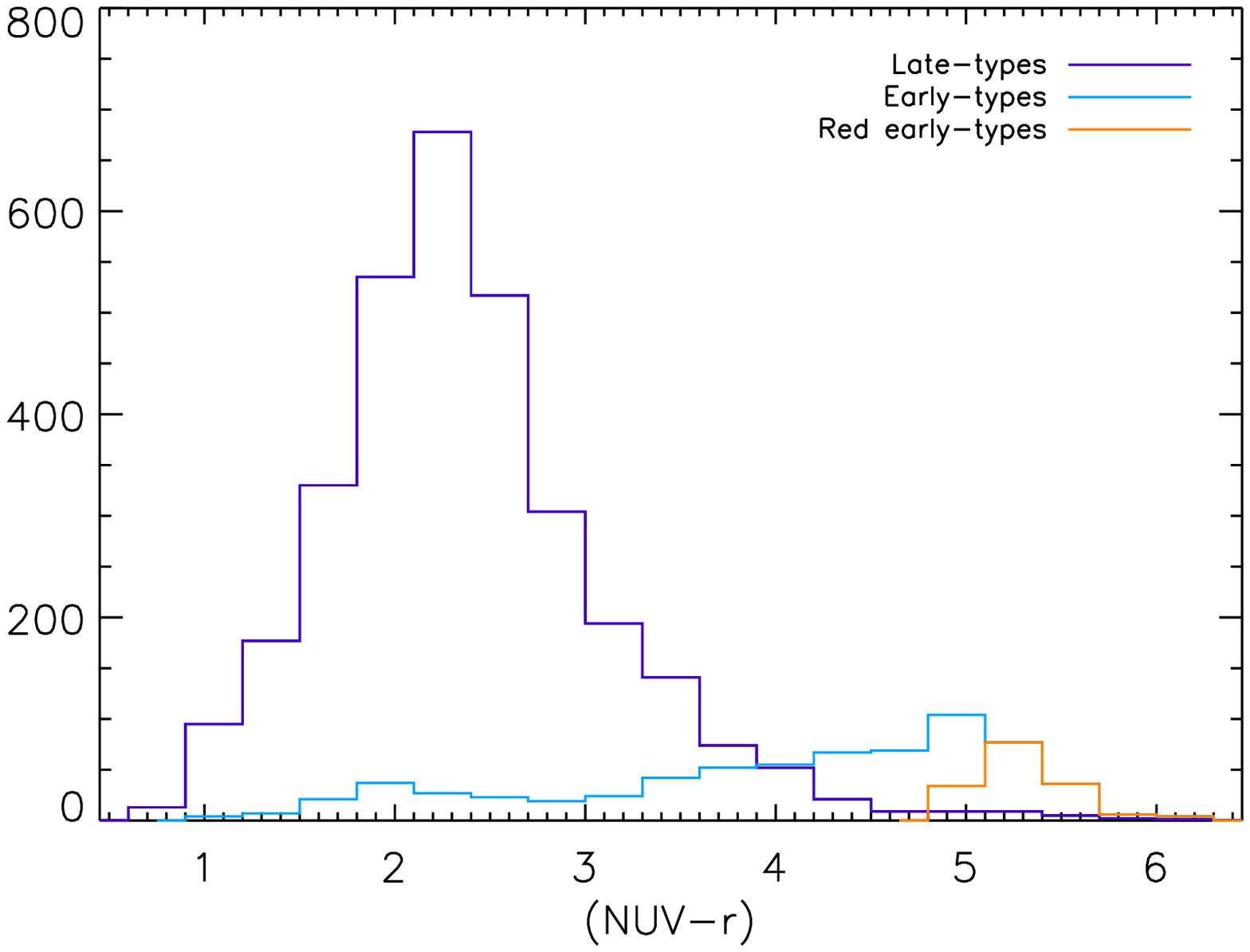}\\
\includegraphics[width=3.5in]{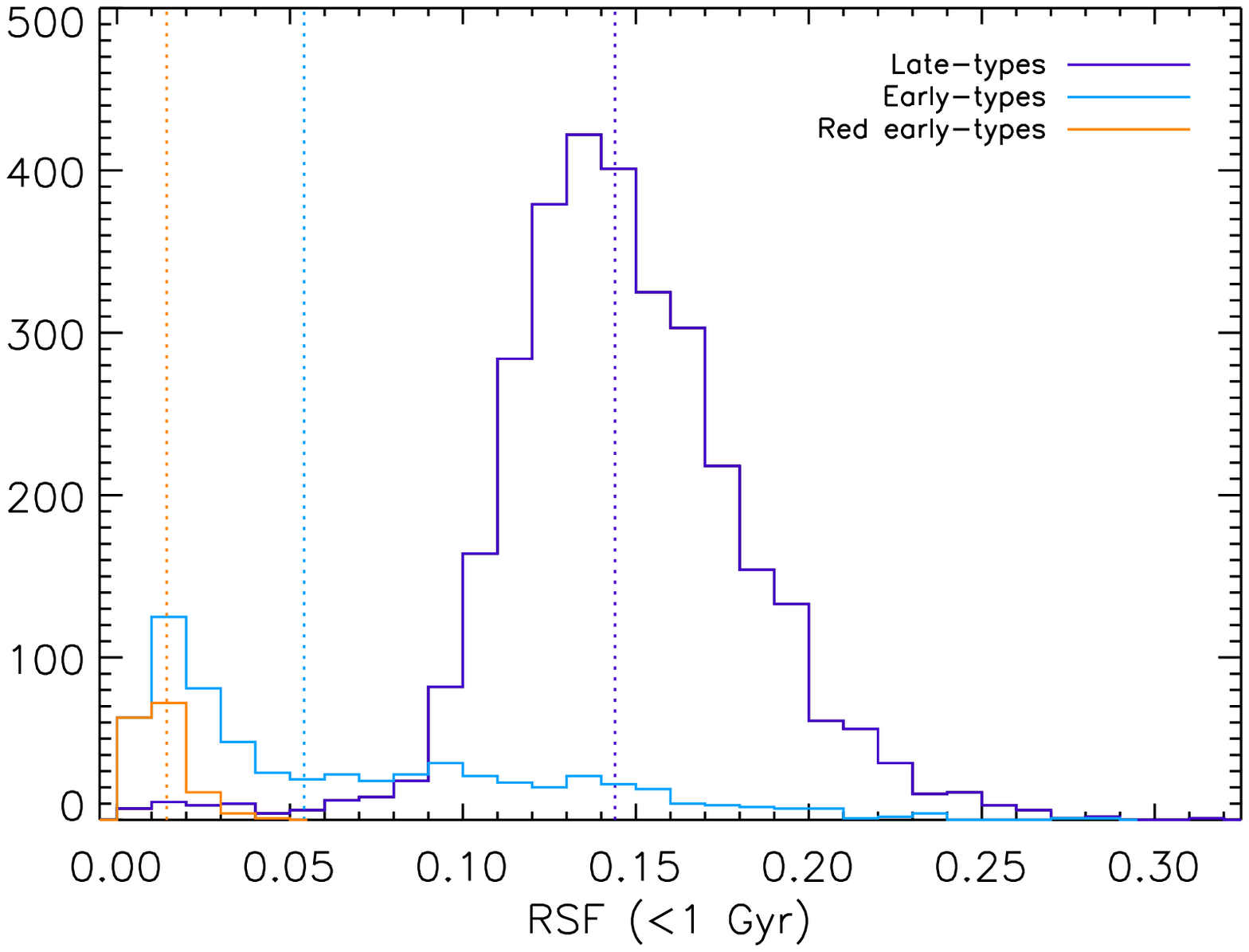}
\end{array}$
\caption{TOP PANEL: Distribution of $(NUV-r)$ colours for
late-types, early-types and red early-type population, defined
using an arbitrary colour cut at $(NUV-r)=5$. BOTTOM PANEL: RSF
distributions for the three categories of objects described above.
Median values are shown using the dotted vertical lines.}
\label{fig:rsf_morphs}
\end{figure}

In Figure \ref{fig:rsf_morphs} (top panel) we show the
distribution of $(NUV-r)$ colours for late-types, early-types and
also the red early-type population. We define `red' early-types
using an arbitrary colour cut at $(NUV-r)=5$. Recall that the red
envelope position for a (dustless) purely old SSP lies at $(NUV-r)
\sim 6.15$ and $(NUV-r) \sim 5.75$ at $z\sim0.5$ and $z\sim1$
respectively. The $(NUV-r)=5$ criterion therefore picks objects
that are less than 1 magnitude away from the red envelope position
in $(NUV-r)$. This selection threshold, although somewhat
arbitrary, is reasonable, given the broadness of the $UV$ red
sequence.

The bottom panel of Figure \ref{fig:rsf_morphs} shows the RSF
distributions for these three categories of objects. The median
value of RSF in the red early-type population is $\sim1.3$
percent. This demonstrates explicitly that the parameter
estimation does not favour dusty SFHs for red early-type systems,
notwithstanding the large range of $E(B-V)$ values available to
it. The spectra of red early-type galaxies are consistent with
them being quiescent systems.

\section{Summary and discussion}
We have combined deep optical and NIR $UBVRIz'JK$ photometry,
photometric redshifts and HST imaging from the MUSYC, COMBO-17 and
GEMS surveys respectively to perform a large-scale study of the
rest-frame $UV$ properties of the early-type galaxy population in
E-CDFS, in the redshift range $0.5<z<1$. Our study is one of the
first of its kind to study the rest-frame $UV$ flux around
$2300\AA$ at high redshift and provides both a useful comparison
to similar studies performed at low redshift and a benchmark for
future work that might exploit the rest-frame $UV$ spectral ranges
to study distant early-type galaxies.

By comparing the observed photometry with synthetic populations,
generated in the framework of the standard model, we have computed
the rest-frame $UV$ and optical colours of the E-CDFS galaxy
population and quantified, for each E-CDFS galaxy, the stellar
mass contributed by recent star formation (RSF), defined as the
mass fraction in stars less than 1 Gyr old in the rest-frame of
the galaxy. The sensitivity of the rest-frame $UV$ to young stars
allows us to detect RSF at levels as low as 1 percent of the total
mass of the galaxy and robustly constrain the star formation
history (SFH) of galaxies within the last Gyr.

We find that, similarly to the low-redshift ($0<z<0.1$) early-type
population, high-redshift early-types display a broad `red
sequence' which spans almost three magnitudes in the $(NUV-r)$
colour. The origin of this broadness is the high sensitivity of
the rest-frame $UV$ to young stars. Small differences in the
proportion of young stars in largely quiescent galaxies broadens
the $UV$ red sequence (bottom panel in Figure \ref{fig:cmr}),
while the optical red sequence remains relatively narrow (top
panel in Figure \ref{fig:cmr}) because optical colours remain
virtually unaffected by small mass fractions of young stars. By
comparing the $(NUV-r)$ colour-magnitude relation of the E-CDFS
galaxy population (top panel in Figure \ref{fig:cmr}) to the
expected $(NUV-r)$ colours of simple stellar populations forming
at high redshift ($z=2,3,5$), we have deduced that $\sim1.1$
percent of early-types in this sample are consistent (within
errors) with \emph{purely} passive ageing since $z=2$. This value
drops to $\sim0.24$ percent and $\sim0.15$ percent for $z=3$ and
$z=5$ respectively.

The early-type population \emph{as a whole} exhibits a typical RSF
between 5 and 13 percent in the redshift range $0.5<z<1$, while
the early-types on the broad red sequence ($NUV-r>4$) typically
show RSF values less than 5 percent. The reddest early-types
(which are also the most luminous) are virtually quiescent with
RSF values of $\sim1$ percent. Within the errors, there is a
slight indication that the RSF increases with redshift, from
$\sim7$ percent at $z=0.5$ to $\sim13$ percent at $z=1$. Note that
the typical uncertainty in the RSF is $\sim2.5$ percent.

Comparison of our results to a corresponding study of early-type
galaxies in the local Universe (Kaviraj et al. 2006b) indicates
that, in contrast to their low-redshift counterparts, the
early-type population in E-CDFS shows a pronounced bimodality in
the $(NUV-r)$ colour distribution around $(NUV-r)\sim3$ (Figure
\ref{fig:comparetolowz}). The peak of the high-redshift $(NUV-r)$
distribution shows a relative shift from its low-redshift
counterpart which is close to what might be expected from the
passive ageing of a simple stellar population forming at high
redshift (indicated by the arrow in Figure
\ref{fig:comparetolowz}). This indicates that the \emph{bulk} of
the stellar population in early-type galaxies, at least on the red
sequence, is overwhelmingly old. The blue peak in the
high-redshift $(NUV-r)$ distribution contains $\sim15$ percent of
the early-type population, with an average RSF of $\sim11$
percent. In comparison, the bluest 15 percent of the low-redshift
early-type population has an average RSF of $\sim6$ percent,
indicating that star formation activity in the most active
early-types has halved between $z=0.7$ and present day.

We note that evolved stellar populations capable of emitting in
the $UV$, e.g. horizontal branch stars, should not be present in
galaxies in this study, since there is insufficient time for them
to appear, given the lower redshift cut employed here ($z=0.5$).
Hence, the $UV$ flux seen in the galaxy population studied here
comes \emph{exclusively} from young stars, making our results more
robust than can be achieved by similar studies at low redshift,
where contamination of the $UV$ spectrum from old populations has
to be taken into account before the contribution from young
populations can be estimated.

This study has demonstrated that the early-type population, in
general, is not composed of galaxies that have been evolving
completely passively since high redshift. Taken together with the
widespread RSF found in low-redshift early-types, we find
compelling evidence that early-type galaxies of all luminosities
form stars over the lifetime of the Universe, although the bulk of
their star formation is already complete at high redshift.
Contrary to conclusions from some past studies, which were
focussed mainly on the optical wavelengths that are insensitive to
recent star formation, this `tail-end' of star formation in
early-type galaxies is measurable and not negligible. Since the
timescale of this study is $\sim2.5$ Gyrs, a simple extrapolation
(from RSF values in Figure \ref{fig:rsf_mv}) indicates that
luminous ($-23<M(V)<-20.5$) early-type galaxies may typically form
up to 10-15 percent of their mass after $z=1$ (with a tail to
higher values), while their less luminous ($M(V)>-20.5$)
counterparts form 30-60 percent of their mass in the redshift
range. These values are probably overestimated since the intensity
of star formation is seen to decrease between $z=0$ and redshifts
probed by this study. This tail-end of star formation should exist
in \emph{intermediate-age} (3-8 Gyr old) stellar populations in
early-type galaxies at present day.

It is worth noting here that the early-type galaxy sample studied
here is drawn from a variety of environments. The largely
photometric nature of the redshifts employed in this study makes
it difficult to separate galaxies in dense regions from their
field counterparts. However, galaxies in dense regions are
predicted to complete their mass assembly and star formation
significantly faster than counterparts in the field (Kaviraj et
al. 2006a). Hence, we should be careful about applying the
conclusions derived above to galaxy populations in general. For
example, although 10 to 15 percent of the stellar mass in luminous
early-types may typically form after $z=1$, the corresponding
values for cluster galaxies will be at least a factor of 2 lower,
given their faster evolution.

We note that our results and, in particular, the implied mass
buildup in the early-type population in the redshift range $0<z<1$
inferred from our analysis, are in good agreement with the results
of \citet{vanderwel2007}, who studied star formation in
high-redshift ($0.85<z<1.15$) early-type galaxies in the CDF-S
using GOODS-MIPS infrared imaging. The majority of their sample is
in the (optical) red sequence (see their Figure 4) and their
results indicate that (a) the early-type population is dominated
by evolved stellar populations (b) most early-types are not
affected by dust-obscured star formation or AGN and (c) the
increase in the early-type stellar mass density in the redshift
range $0<z<1$ is $14\pm7$ percent. (a), (b) and (c) are clearly
consistent with our results for early-types on the $UV$ red
sequence and (d) compares well with our estimate of 10-15 percent
for the mass buildup expected for luminous early-types which
inhabit the red sequence.

While the analysis presented in this study has quantified the RSF
in the E-CDFS early-type population, we have not explored the
possible sources of young stars in these galaxies. Although a
comprehensive study of these sources is beyond the scope of this
paper (and will be dealt with in future work), we use the
results of our analysis to speculate on the RSF channels that
might be at work in the early-type population.

Several plausible sources of low-level star formation exist, which
could, either individually or collectively, explain the broadness
of the $UV$ red sequence. An obvious source of star formation is
internally recycled gas from stellar mass loss. For standard IMFs,
a substantial (20-30 percent) fraction of the stellar mass may be
returned to the ISM. However, only stars formed from recycled gas
entering the ISM within the last Gyr will contribute to the $UV$.
Kaviraj et al. (2006c; see their Section 6) have suggested that
RSF from internal mass loss is an order of magnitude too small
($\sim 0.2$ percent; see their Figure 28, bottom left panel) than
what is observed (a few percent) in red early-types in this study.
It is therefore likely that internal mass loss plays a negligible
role in broadening the red sequence and that other sources of gas
are required to produce the observed RSF in these systems.

Condensation from the extensive hot gas reservoirs hosted by
massive early-type galaxies may provide an additional source of
cold gas in these systems. While feedback sources (e.g. AGN and
supernovae) might be expected to maintain the temperature of the
hot gas reservoir and evaporate infalling cold material in the
most massive haloes \citep{Binney2004}, this process may not be
fully efficient, allowing some gas condensation to take place,
which could then result in low-level star formation as the cold
gas settles in the potential well.

Merger and accretion events provide an alternative source of young
stars. The small levels of RSF in red sequence early-types
indicate that the star formation in these galaxies could be driven
either by accretion of small gas-rich satellites or through
\emph{largely} dry equal mass mergers. Simulations of binary
mergers at low redshift (Peirani, Kaviraj et al., in prep)
indicate that galaxies on the broad $UV$ red sequence can be
reproduced by `minor' mergers that have mass ratios between 1:6
and 1:10 and where the accreted satellites have high gas fractions
($>20$ percent). While it is difficult to discriminate between
gas-rich satellite accretion and largely dry major mergers based
on the RSF values alone (since both scenarios can supply equal
amounts of gas), the favoured mechanism for RSF in the red
sequence could possibly be deduced by studying the rates of minor
and \emph{red} major mergers at the redshifts sampled by this
study. In a recent work, \citet{Bell2006} found that 12 out of 379
galaxies ($\sim3$ percent) in the E-CDFS \emph{optical} red
sequence are involved in major mergers (mass ratios greater than
1:4) and deduced that luminous ($M(V)<-20.5$) early-types may have
undergone 0.5-2 major dry mergers since $z\sim0.7$. It therefore
seems unlikely that major mergers alone can account for virtually
the entire $UV$ red sequence hosting low level RSF. Minor mergers
(which are predicted to be more frequent in the hierarchical
model) should therefore play a significant part in broadening the
red sequence.

Early-types residing in the blue cloud necessarily require a more
gas-rich formation scenario and a likely mechanism for creating
these objects are gas-rich equal mass mergers. It is worth noting
here that galaxies in the `possible early-type' category almost
exclusively inhabit the blue cloud (see Figure \ref{fig:cmr}) and
some appear to show asymmetries in their structure that may be
indicative of a recent interaction (see the examples in Figure
\ref{fig:petype_examples}). It is reasonable to speculate that,
given their largely spheroidal appearance and blue colours, some
of the `possible early-types' may be precursors of the relaxed
early-type population in the blue cloud, implying that these blue
early-types may also have had gas-rich interactions in the recent
past. It is possible that the morphological signatures of the
interactions are not visible in the images but the high stellar
mass fraction formed in the burst imparts the (spheroidal) remnant
with its observed blue $UV$ colour. Finally, it is worth noting
that if the vigorous star formation episode in these galaxies is
quenched by e.g. SN or AGN feedback (Kaviraj et al. 2007), the
galaxies will naturally evolve onto the red sequence. Therefore a
small fraction of the broad red sequence could consist of such
galaxies in the process of migration. However, since the migration
is towards both redder colours and fainter magnitudes, this would
affect a negligible part of the red sequence in this study due to
the lack of red galaxies which have $m(R)>24$ in this sample (see
bottom panel of Figure 3).

In summary, it seems reasonable to speculate that the broadness of
the red sequence is driven by a combination of condensation from
the hot gas reservoirs coupled with accretion of gas-rich
satellites (and to a lesser extent dry equal mass mergers), while
early-types on the blue cloud are formed through gas-rich equal
mass mergers that impart the remant with its spheroidal morphology
and leads to the high observed values of RSF.

The RSF values derived in this study have some interesting
implications for the alpha-enhancement ($[\alpha/Fe]$) observed in
the spectra of nearby early-type galaxies. Massive early-types at
low redshift frequently exhibit super-solar values of
alpha-enhancement ($[\alpha/Fe]\sim 0.3$ dex), which indicates
that the bulk of the star formation in these galaxies takes place
over a timescale shorter than the typical onset time of Type Ia
supernovae (SN). While alpha elements (such as $Mg$) are provided
by Type II SN that appear on very short timescales of $\sim10^6$
yrs, the bulk of the $Fe$ is supplied by Type Ia ejecta which are
expected to appear after typical timelags of $\sim 1$ Gyr. Hence,
if star formation takes place over timescales greater than a Gyr,
subsequent generations of stars incorporate the $Fe$ injected into
the inter-stellar medium, resulting in a decrease in the
alpha-enhancement and $[\alpha/Fe]$ values that reflect those in
the solar neighbourhood.

It is instructive to estimate how much dilution we would expect in
the early-type population from the RSF values derived in this
study. If we assume that (a) the bulk stellar mass forming at high
redshift is indeed created over a short timescale and has an
alpha-enhancement characteristic of monolithic collapse
($[\alpha/Fe]\sim 0.3$ dex) and (b) the rest of the mass forms
over a longer timescale, at lower redshifts ($z<1$) and has an
alpha-enhancement characterstic of the solar neighbourhood, then
the mass-weighted $[\alpha/Fe]$ at $z\sim0$ can be expressed
approximately as

\begin{equation}
[\alpha/Fe] \sim (1-f).[\alpha/Fe]_{ML} + f.[\alpha/Fe]_{\odot},
\end{equation}

where $f$ is the total mass fraction of stars formed at low
redshifts ($z<1$), $[\alpha/Fe]_{ML}$ is the alpha-enhancement of
a `monolithic collapse' type starburst and $[\alpha/Fe]_{\odot}$
is the alpha-enhancement characteristic of the solar
neighbourhood.

A strict monolithic collapse would create a stellar population
with $[\alpha/Fe]\sim 0.3$ dex (see e.g. top right panel of
Kaviraj et al. 2006c). Since the observed $[\alpha/Fe]$ values in
early-type systems typically tend to be in excess of 0.2
\citep[e.g][]{Thomas1999}, Eqn (1) indicates that, with the
stellar mass fractions expected to form in the bulk of the
early-type population after $z\sim1$ ($<13$ percent), the inferred
alpha-enhancements in these galaxies should not deviate
significantly from the monolithic value. While Eqn (1) is
approximate, the RSF values we have derived in this study are in
agreement with the observed alpha-enhancements of the local
early-type galaxy population.

The sensitivity of the rest-frame $UV$ spectrum to the recent SFH
allows us sample the build-up of the SFH curve of galaxy
populations over time. Clearly, access to the rest-frame $UV$ over
a large redshift range allows us to reconstruct the SFH curve with
a reasonable degree of accuracy. While this study, together with
similar work at low redshift (Kaviraj et al. 2006b), are the first
attempts at such a reconstruction, future studies using the
rest-frame $UV$ are keenly anticipated, especially at higher
redshifts ($z>2$) where massive early-type galaxies are in the
process of or have just completed their major episodes of star
formation and assembly.


\nocite{Kaviraj2006a} \nocite{Kaviraj2006b} \nocite{Kaviraj2006c}
\nocite{SDSSDR4} \nocite{Gawiser2006} \nocite{Rix2004}
\nocite{Lefevre2004} \nocite{Giavalisco2004} \nocite{Kaviraj2007}
\nocite{Bernardi2003a} \nocite{Martin2005} \nocite{WMAP_3}


\section*{Acknowledgements}
We are grateful to Roger Davies and Rachel Somerville for numerous
discussions which improved the quality of this work. Seong-Hee Kim
is thanked for generously providing sample metallicity
distribution functions from his galaxy formation models. Rob
Kennicutt, Caina Hao, Chris Wolf, Claudia Maraston and Daniel
Thomas are thanked for useful comments. SK acknowledges a
Leverhulme Early-Career Fellowship, a BIPAC fellowship and a
Junior Research Fellowship from Worcester College, University of
Oxford.


\bibliographystyle{mn2e}
\bibliography{references}


\end{document}